\documentclass[12pt,preprint]{aastex}

\newcommand{\swf}{{\it Swift}}
\newcommand{\fer}{{\it Fermi}}
\newcommand{\sax}{{\it BeppoSAX}}

\newcommand{\xmm}{{\it XMM-Newton}}

\usepackage{graphicx}
\usepackage{longtable}

\slugcomment{version \today: fm}

\shorttitle{X-ray spectral curvature of HBLs}
\shortauthors{F. Massaro, A. Paggi, M. Elvis, A. Cavaliere 2011}

\begin{document}
\title{X-ray spectral curvature of High Frequency Peaked BL Lacs: \\ a predictor for the TeV flux}
\author{F. Massaro\altaffilmark{1}, A. Paggi\altaffilmark{1,2}, M. Elvis\altaffilmark{1}, A. Cavaliere\altaffilmark{2}
}

\affil{Harvard - Smithsonian Astrophysical Observatory, 60 Garden Street, Cambridge, MA 02138}
\affil{Dipartimento di Fisica, Universit\`{a} di Roma Tor Vergata, Via della Ricerca scientifica 1, I-00133 Roma, Italy}

\begin{abstract} 
Most of the extragalactic sources detected at TeV energies are BL Lac objects.
They belong to the subclass of ``high frequency peaked BL Lacs" (HBLs) exhibiting spectral
energy distributions with a lower energy peak in the X-ray band; this is widely interpreted as
synchrotron emission from relativistic electrons. The X-ray spectra are generally curved, and well described in terms of a log-parabolic shape.
In a previous investigation of TeV HBLs (TBLs) we found two correlations between their spectral parameters.
(1) The synchrotron peak luminosity $L_p$ increases with its peak energy $E_p$; (2) the curvature parameter $b$ 
decreases as $E_p$ increases. The first is consistent with the
synchrotron scenario, while the second is expected from statistical/stochastic acceleration mechanisms for the emitting electrons.
Here we present an extensive X-ray analysis of a sample of HBLs observed with \xmm~and \swf~but  undetected at TeV energies (UBLs), 
to compare their spectral behavior with that of TBLs. 
Investigating the distributions of their spectral parameters and
comparing the TBL X-ray spectra with that of UBLs, we develop
a criterion to select the best HBLs candidates for future TeV observations.
\end{abstract}


\section{Introduction}
The great majority ($\geq$ 80\%) of the extragalactic sources detected to April 2011 in $\gamma$ rays at TeV energies are BL Lac objects.
These are a class of active galactic nuclei (AGNs) characterized by strong 
and highly variable non-thermal radiations from radio frequencies to TeV energies.
Their observational properties include weak or absent emission lines, two-hump shaped 
spectral energy distributions (SEDs, i.e., log $\nu F_{\nu}$ $vs$ log $\nu$), high radio and optical polarization,
and superluminal motions. These are interpreted as the result of radiation from a relativistic jet closely aligned to the 
line of sight \citep{blandford78}.

BL Lacs come in two flavors: the ``high-frequency peaked BL Lacs" (HBLs)
in which the low energy component of the SED peaks between the UV band and 
X-rays, and the ``low-frequency peaked BL Lacs" (LBLs)
when the SED peak falls in the IR-optical range \citep{padovani95}.
It is widely agreed that this low-energy 
component is produced by synchrotron radiation 
of ultrarelativistic particles (i.e., electrons) accelerated in the jets, while
the high energy component is likely due to inverse-Compton
scattering of the synchrotron photons by the same electron population 
(Synchrotron Self-Compton, SSC, see e.g. \citealt{marscher85,inoue96}).

In the following, we distinguish the HBLs detected at TeV energies from those not yet detected; 
we refer to the former as TBLs, and to the latter as UBLs.

A useful phenomenological description of the BL Lac X-ray spectra was introduced by \citet{landau86} in terms 
of a {\it log-parabolic} (LP) model (i.e., a  parabolic shape in a 
double-log plot); subsequently, this model has been frequently adopted
for the low energy bump, e.g., by \citet{tanihata04}, \citet{massaro04} and other authors.
Recently, the high energy component at TeV energies has also been 
successfully modeled with the same spectral shape \citep{massaro06,aharonian09,aleksic11,acciari11,abdo11}.
We note that such LP synchrotron spectra are emitted by log-parabolic particle energy distributions (PEDs), 
obtained via the Fokker-Planck equation from a mono-energetic electron 
injection subjected to systematic and stochastic accelerations \citep{kardashev62,massaro06,stawarz08,paggi09}.

The LP model has been used also to describe the SED of 
other classes of jet-dominated sources: plerions \citep{campana09}, high frequency
peaked (HFPs) radio sources \citep{maselli09}, and, recently, 
Solar Flares\citep{grigis08} and Gamma-Ray Bursts (GRBs) \citep{massaro10a,massaro11a}.

Adopting the LP model, the X-ray SED of HBLs is described in terms of 3 parameters: 
(1) the peak energy, $E_p$, in $\nu F_{\nu}$ space, (2) the maximum height of the SED, $S_p$, 
evaluated at $E_p$ (or the corresponding peak luminosity 
$L_p$ $\simeq$ $4\pi D^2_L S_p$, with $D_L$ being the luminosity distance), 
and (3) the spectral curvature, $b$, around $E_p$
(\citealt{tramacere07}, Massaro et al. 2008a, hereafter \citeauthor{massaro08a}).

Extensive investigations of the TBLs,  
based on all the X-ray observations available in the \sax, \xmm~and \swf~archives 
between 1997 and 2007, have shown that several TBLs
trace two correlations in the ($E_p, L_p, b$) parameter space: 
(1) the peak luminosity $L_p$ 
increases with $E_p$, as expected in the synchrotron scenario,
(2) the curvature parameter $b$ decreases as $E_p$ increases (\citeauthor{massaro08a})
as expected in a stochastic acceleration scenario (e.g., \citealt{tramacere07}).

As a result, TBLs cover a well-constrained region in  the $E_p-b$ plane (hereinafter the ``acceleration plane").
The correlation between $b$ and $E_p$ is evident for the 16 TBLs in \citeauthor{massaro08a},
whilst no clear trend in the $E_p-L_p$ plane has been found for the whole sample.

Many HBLs have been targeted at TeV energies by HESS, Magic and VERITAS, 
but by no means all of them have been detected. 
It is striking that 19 out of 24 TBLs (to 2010, August 1st) 
belong to the Einstein Slew Survey Sample of BL Lacertae Objects (1ES, \citealt{elvis92,perlman96}),
which includes only the brightest X-ray extragalactic sources at $\sim$ 1 keV. The remaining TBLs
belong to three different samples, namely: 
1) The ROSAT All-Sky Survey-Green Bank BL Lac catalog (RGB, \citealt{laurent99});
2) The sedentary survey of extreme high energy peaked BL Lacs 
(SHBL\footnote{\underline{http://www.asdc.asi.it/sedentary/}}, \citealt{giommi05});
3) The Hubble Space Telescope Survey of BL Lacertae Objects;
(HST, \citealt{scarpa99,urry00}) (see Table \ref{tabella1}).
Consequently, we selected all the UBLs in the above four samples 
to search for possible differences between these sources and the TBLs.

In this paper, we present the sample selection criteria, the data reduction and data analysis procedures
adopted to perform our investigation.
Finally, comparing the distribution of the X-ray spectral parameters, we define criteria 
to predict future TBLs on the basis of X-ray observations only.
The theoretical aspects and the interpretation of the observational results will be presented in 
\citet{massaro11b}. 

We use cgs units unless stated otherwise
and we assume a flat cosmology with $H_0=72$ km s$^{-1}$ Mpc$^{-1}$,
$\Omega_{M}=0.26$ and $\Omega_{\Lambda}=0.74$ \citep{dunkley09}.

\section{Sample selection}
We chose all the sources classified as BL Lac objects or BL Lac candidates in the ROMA BZCAT
\footnote{\underline{http://www.asdc.asi.it/bzcat/}} \citep{massaro09,massaro10b}
that are present in the four samples in which TBLs are found (see Section 1), excluding the TBLs.

To compare the behavior of TBLs and UBLs, we selected a sample of UBLs on adopting the following criteria.
\begin{itemize}
\item{We calculated the ratio $\Phi_{XR}$ between the X-ray flux $F_X$ (0.1 - 2.4 keV) and the radio flux $S_{1.4}$ (at 1.4 GHz), $\Phi_{XR}$ 
(i.e., $\Phi_{XR} = 10^{-3}\,F_X/(S_{1.4}\,\Delta \nu)$ erg\,cm$^{-2}$\,s$^{-1}$Jy$^{-1}$ with $\Delta \nu$ = 1GHz),
using the values of $F_X$ and $S_{1.4}$ reported in the ROMA BZCAT \citep{massaro09,massaro10b}.
We select BL Lacs with $\Phi_{XR}$ $\geq$ 0.1 that corresponds to HBLs, according to the criterion established by \citet{maselli09}.}
\item{We restricted our sample to those sources with redshift $z \leq 0.539$, the highest redshift for an 
extragalactic TeV source (i.e., 3C~279, see Albert et al. 2008).
Using this cut in redshift, we assumed that any extragalactic source with $z \geq 0.539$ could not be detected at TeV energies,
because of the absorption by the extragalactic background light \citep{dwek05}.}
\item{We considered only UBLs with X-ray observations, up to the end of October 2010, in the \xmm~or \swf~archives, as performed for the TBLs by \citeauthor{massaro08a}
that have an exposure longer than 150 s, in order to have a good chance of detection and a sufficient number of 
counts to perform the X-ray spectral analysis (see also \citeauthor{massaro08a}).}
\end{itemize}

There are 118 UBLs with known redshift. in the four samples considered.
However, 71 UBLs are excluded by requiring $\Phi_{XR} \geq 0.1$, $z \leq$ 0.539 and with X-ray observations with exposure longer than 150 s.
The remaining 47 UBLs constitute the sample we analyze below.

These 47 UBLs a total of 135 X-ray observations: 123 \swf~observations and 12 by \xmm.
Only 19 UBLs out of the total 47 selected targets have been detected
by \fer~during the first year of operations \citep{abdo10}.

%-----------------------------------------------------------------------------------------------------------------------
\begin{table}
\caption{The properties of the BL Lac samples.}\label{tabella1}
\begin{center}
\begin{tabular}{|cccccc|}
\hline
(1)  & (2) & (3) & (4) & (5) & (6) \\
Sample  & z$_{max}$ & Total & TBLs & HBLs & UBLs \\
\hline
\noalign{\smallskip}
1ES    & 0.940 &  55 & 18 &  46 &  7 \\
HST    & 0.940 &  94 & 19 &  57 &  3 \\
SHBL   & 0.702 & 122 &  9 & 122 & 29 \\
RGB    & 0.664 & 109 &  7 &  70 &  7 \\
\noalign{\smallskip}
\hline
\end{tabular}\\
\end{center}
Col.(2) Total number of BL Lacs in the sample. 
Col.(3) Highest redshift in the sample. 
Col.(4) Number of TBLs present in the sample. 
Col.(5) Number of HBLs in the sample.
Col.(6) Number of UBLs selected.
\end{table}
%-----------------------------------------------------------------------------------------------------------------------
Table \ref{tabella1} reports:
the highest redshift for the sample (Col. 2),
the number of BL Lacs identified in the ROMA BZCAT (Col. 3),
the number of TBLs in the sample (Col. 4), 
the number of HBLs present (Col. 5) 
and the UBLs selected according to the criteria defined above (Col. 6). 

The basic data for all the 47 selected UBLs are reported in Table \ref{tabella2}:
the ROMA BZCAT name (Col. 1) and sample name (Col. 2), the equatorial coordinates (J2000) (Col. 3 and Col. 4),
the redshift (Col. 5, from \citealt{massaro10b}), the luminosity distance $D_L$ (Col. 6), the value of the Galactic column density
$N_{H,Gal}$ (Col. 7, see \citealt{kalberla05}), the X-ray to radio flux ratio $\Phi_{XR}$ (Col. 8) and the number of both the
\xmm~and \swf~observations (Col. 9 and Col. 10 respectively).
Finally, in Col. (11) we show the TeV candidate class provided by our investigation discussed in Section 6.

%Finally, we did not analyze two NBLs: BZB~J0333-3619 and BZB~J0613+7107.
%These sources are in the FOV of two deeply observed Seyfert galaxies: NGC 1365 and Mrk 3, respectively.
%A separate paper including their \chn~observations is in preparation.
%------------------------------------------------------------------------------------------------------------------------------------------------------------------------------
\begin{table*}
\caption{UBLs selected.}\label{tabella2}
\begin{center}
\tiny
\begin{tabular}{|cccccccccccc|}
\hline
(1) & (2) & (3) & (4) & (5) & (6) & (7) & (8) & (9) & (10) & (11) & (12) \\
BZCAT~Name & Other~Name  & RA    & DEC   & z & D$_L$ & N$_{H,Gal}$       &$\Phi_{XR}$& \swf &{\it XMM}& \fer & TeV \\
           &             &(J2000)&(J2000)&   & [Mpc] &$[10^{20}cm^{-2}]$ &           &      &         &      &class\\
\hline
\noalign{\smallskip}
BZB~J0013-1854    &1RXS~J001356.6-18540 & 00 13 56.0 & -18 54 06.0 & 0.094 &  420.1  & 2.13 & 2.24 & 4   & --- & --- & 3 \\ % new Feb 2011 - candidate
BZB~J0123+3420    & 1ES~0120+340        & 01 23 08.5 & +34 20 47.0 & 0.272 & 1359.7  & 5.20 & 5.74 & 17  &  1  & --- & 3 \\ 
BZB~J0201+0034    & 1ES~0158+003        & 02 01 06.1 & +00 34 00.0 & 0.298 & 1511.2  & 2.23 & 2.71 & 1   & --- & --- & - \\    
BZB~J0208+3523    &1RXS~J020837.5+35231 & 02 08 38.2 & +35 23 13.0 & 0.318 & 1629.9  & 6.27 & 5.76 & --- &  2  &  y  & 2 \\
BZB~J0214+5144    & RGB~J0214+517       & 02 14 17.8 & +51 44 52.0 & 0.049 &  212.0  & 14.4 & 0.16 & 3   & --- & --- & 3 \\
BZB~J0227+0202    &1RXS J022716.6+02015 & 02 27 16.5 & +02 02 00.0 & 0.456 & 2499.3  & 2.67 & 5.05 & 2   & --- & --- & - \\ % new Feb 2011 
BZB~J0325-1646    &1RXS~J032540.8-16460 & 03 25 41.1 & -16 46 14.9 & 0.291 & 1470.1  & 3.27 & 10.1 & 3   & --- &  y  & - \\ % candidate
BZB~J0326+0225    & 1ES~0323+022        & 03 26 13.9 & +02 25 14.0 & 0.147 &  681.2  & 7.87 & 1.77 & 3   &  1  &  y  & 1 \\
BZB~J0441+1504    &1RXS~J041112.1-39413 & 04 41 27.4 & +15 04 54.0 & 0.109 &  492.3  & 14.0 & 7.28 & 1   &  1  & --- & - \\
BZB~J0442-0018    &1RXS~J044229.8-00182 & 04 42 29.8 & -00 18 34.9 & 0.449 & 2453.2  & 4.83 & 3.35 & 4   & --- &  y  & 1 \\
BZB~J0621-3411    &1RXS~J062150.0-34114 & 06 21 49.4 & -34 11 53.9 & 0.529 & 2991.5  & 4.08 & 2.30 & 1   & --- & --- & - \\
BZB~J0744+7433    & 1ES~0737+746        & 07 44 05.2 & +74 33 56.9 & 0.314 & 1606.0  & 3.28 & 2.74 & --- &  2  &  y  & 1 \\
BZB~J0751+1730    &1RXS~J075124.3+17304 & 07 51 25.0 & +17 30 51.0 & 0.185 &  878.4  & 4.93 & 1.78 & 1   & --- & --- & - \\
BZB~J0753+2921    &1RXS~J075322.4+29215 & 07 53 24.6 & +29 21 31.0 & 0.161 &  752.9  & 3.44 & 0.28 & 1   & --- & --- & - \\ % new Feb 2011
BZB~J0847+1133    &1RXS~J084713.3+11334 & 08 47 12.8 & +11 33 50.0 & 0.199 &  953.2  & 3.17 & 3.34 & 1   & --- &  y  & - \\
BZB~J0916+5238    & RGB~J0916+526       & 09 16 51.8 & +52 38 27.9 & 0.190 &  905.0  & 1.43 & 0.53 & 1   & --- & --- & - \\
BZB~J0930+4950    &1RXS~J093037.1+49502 & 09 30 37.5 & +49 50 25.0 & 0.187 &  889.1  & 1.38 & 7.94 & 1   & --- & --- & - \\ % new Feb 2011
BZB~J0952+7502    &1RXS~J095225.8+75021 & 09 52 24.1 & +75 02 12.9 & 0.179 &  846.8  & 2.23 & 2.07 & 2   & --- & --- & - \\
BZB~J1010-3119    &1RXS~J101015.9-31190 & 10 10 15.9 & -31 19 08.0 & 0.143 &  660.9  & 8.48 & 1.37 & 2   & --- & --- & 3 \\
BZB~J1022+5124    &1RXS~J102212.5+51240 & 10 22 12.6 & +51 23 59.9 & 0.142 &  655.9  & 1.02 & 6.88 & 1   & --- & --- & - \\
BZB~J1053+4929    & RGB~J1053+494       & 10 53 44.0 & +49 29 56.0 & 0.140 &  645.8  & 1.50 & 0.13 & 1   & --- &  y  & - \\ 
BZB~J1056+0252    &1RXS~J105607.0+02521 & 10 56 06.6 & +02 52 13.0 & 0.236 & 1155.7  & 3.82 & 17.3 & 1   & --- & --- & - \\
BZB~J1111+3452    &1RXS~J111131.2+34521 & 11 11 30.7 & +34 52 02.9 & 0.212 & 1023.5  & 1.64 & 5.71 & 1   & --- & --- & - \\ % candidate
BZB~J1117+2014    &1RXS~J111706.3+20141 & 11 17 06.1 & +20 14 08.0 & 0.139 &  640.7  & 1.35 & 3.26 & 1   & --- &  y  & - \\ 
BZB~J1136+6737    & 1136+676            & 11 36 29.9 & +67 37 04.0 & 0.136 &  625.7  & 1.09 & 3.28 & 5   & --- &  y  & 2 \\
BZB~J1145-0340    &1RXS~J114535.8-03394 & 11 45 35.1 & -03 40 00.9 & 0.167 &  784.0  & 2.22 & 2.28 & 2   & --- & --- & - \\
BZB~J1154-0010    &1RXS~J115404.9-00100 & 11 54 04.5 & -00 10 09.0 & 0.254 & 1256.9  & 2.06 & 2.75 & 1   & --- & --- & - \\
BZB~J1231+6414    & 1229+643            & 12 31 31.3 & +64 14 17.9 & 0.163 &  763.3  & 2.12 & 0.43 & --- &  1  & --- & - \\
BZB~J1237+6258    &1RXS~J123739.2+62584 & 12 37 38.9 & +62 58 41.9 & 0.297 & 1505.3  & 0.97 & 1.90 & 13  &  2  & --- & - \\
BZB~J1253-3931    &1RXS~J125341.2-39320 & 12 53 41.2 & -39 31 59.0 & 0.179 &  846.8  & 7.66 & 1.47 & 1   & --- & --- & 3 \\ 
BZB~J1257+2412    & 1ES~1255+244        & 12 57 31.9 & +24 12 39.9 & 0.141 &  650.8  & 1.25 & 5.16 & 1   &  1  & --- & - \\
BZB~J1341+3959    & RGB~J1341+399       & 13 41 05.1 & +39 59 44.9 & 0.172 &  810.0  & 0.80 & 1.26 & 4   & --- &  y  & - \\
BZB~J1417+2543    &1RXS~J141756.8+25432 & 14 17 56.5 & +25 43 26.0 & 0.237 & 1161.3  & 1.54 & 1.72 & 5   & --- &  y  & 2 \\
BZB~J1439+3932    &1RXS~J143917.7+39324 & 14 39 17.5 & +39 32 42.0 & 0.344 & 1787.1  & 1.14 & 2.64 & 2   & --- &  y  & - \\
BZB~J1442+1200    & 1ES~1440+122        & 14 42 48.1 & +12 00 39.9 & 0.163 &  763.3  & 1.58 & 1.13 & 4   & --- &  y  & 2 \\
BZB~J1510+3335    &1RXS~J151040.8+33351 & 15 10 41.1 & +33 35 04.0 & 0.114 &  516.7  & 1.54 & 3.16 & --- &  1  & --- & - \\ % new Feb 2011
BZB~J1534+3715    & RGB~J1534+372       & 15 34 47.2 & +37 15 54.0 & 0.143 &  660.9  & 1.33 & 0.10 & 1   & --- & --- & - \\
BZB~J1605+5421    &1RXS~J160518.5+54210 & 16 05 19.0 & +54 21 00.0 & 0.212 & 1023.5  & 0.89 & 5.53 & 1   & --- & --- & - \\
BZB~J1626+3513    & RGB~J1626+352       & 16 26 25.8 & +35 13 41.0 & 0.497 & 2773.1  & 1.36 & 0.35 & --- &  2  & --- & - \\ % new Feb 2011
BZB~J1728+5013    & 1728+502            & 17 28 18.5 & +50 13 09.9 & 0.055 &  239.0  & 2.35 & 1.01 & 4   & --- &  y  & 2 \\
BZB~J1743+1935    & 1ES~1741+196        & 17 43 57.7 & +19 35 08.9 & 0.080 &  354.0  & 7.36 & 0.14 & 3   & --- &  y  & 1 \\
BZB~J2131-0915    &1RXS~J213135.5-09152 & 21 31 35.3 & -09 15 21.9 & 0.449 & 2453.2  & 3.62 & 1.74 & 1   & --- &  y  & - \\
BZB~J2201-1707    &1RXS~J220156.0-17065 & 22 01 55.8 & -17 07 00.0 & 0.169 &  794.4  & 2.91 & 5.92 & 2   & --- & --- & - \\ % new Feb 2011 - candidate
BZB~J2250+3824    & RGB~J2250+384       & 22 50 05.7 & +38 24 37.0 & 0.119 &  541.2  & 10.4 & 0.24 & 16  & --- &  y  & 2 \\
BZB~J2308-2219    &1RXS~J230846.7-22195 & 23 08 46.8 & -22 19 49.0 & 0.137 &  630.7  & 1.86 & 7.12 & 1   & --- & --- & - \\
BZB~J2322+3436    & RGB~J2322+346       & 23 22 43.9 & +34 36 14.0 & 0.098 &  439.3  & 6.83 & 0.11 & 2   & --- &  y  & - \\
BZB~J2343+3439    &1RXS~J234332.5+34395 & 23 43 33.5 & +34 39 48.9 & 0.366 & 1922.6  & 6.75 & 1.60 & 2   & --- &  y  & - \\
\noalign{\smallskip}
\hline
\end{tabular}\\
\end{center}
Col. (1) ROMA BZCAT source names. Col. (2) the name in the selected sample. Cols.(3,4) the right ascension and declination, respectively. 
Col. (4) gives the redshift (from ROMA BZCAT). Col. (5) reports the luminosity distance. Cols. (6) the Galactic column density along the line of sight 
\citep{kalberla05}. Col. (8) the X-ray to radio flux ratio $\Phi_{XR}$ (see Section 2). Cols. (9,10) report the number of X-ray observations per satellite. 
Col. (11) indicates if the source has been detected in the \fer~LAT 1st year catalog, wile Col. (12) the TeV candidate class derived from our analysis (see Section 6) 
\end{table*}
%------------------------------------------------------------------------------------------------------------------------------------------------------------------------------

\section{Data reduction procedures}
The reduction procedure for the \xmm~data follows that described in \citet{tramacere07}; 
additional details on both the \xmm~and \swf~data reduction procedures 
can be found in \citeauthor{massaro08a} and \citet{massaro08b}.
In the following subsections we report only the basic details. 

\subsection{\xmm~observations}
The sources were observed with \xmm~by means of
all EPIC CCD cameras: the EPIC-PN \citep{struder01},  and EPIC-MOS \citep{turner01}. 
%We consider only EPIC-MOS data for our analyses.

Extractions of light curves, source  and background spectra were done with
the \xmm~ Science  Analysis System (SAS) v6.5.0.  The Calibration
Index File  (CIF) and  the summary file  of the Observation  Data File
(ODF) were  generated using Updated Calibration Files (CCF) following the ``User's Guide  to
the \xmm~ Science Analysis System" (issue 3.1, \citealt{loiseau04}) and ``The \xmm~
ABC Guide"  (vers. 2.01, \citealt{snowden04}). 
Event files were produced by the EMCHAIN pipeline.

Light curves for  each dataset were extracted,  and  all high-background intervals 
filtered out to exclude  time intervals contaminated  by  solar  flares. 
Then, by visual inspection, we selected good time intervals (GTI) far from solar flare peaks that have
no count rate variations on time scales shorter than 500 seconds.
Photons are extracted from an annular region using
different apertures to minimize  pile-up, which affects MOS data.  The
mean value of the external radius used for the annular region is $40$~$\prime\prime$.

A slightly restricted energy range (0.5--10 keV) is used to minimize
residual calibration uncertainties.   
To ensure the validity of Gaussian  statistics,  data have been grouped  by  combining
instrumental channels so that each bin contains 30 counts or more.
%-----------------------------------------------------------------------------------------------------------------------
\begin{figure*}[!htp]
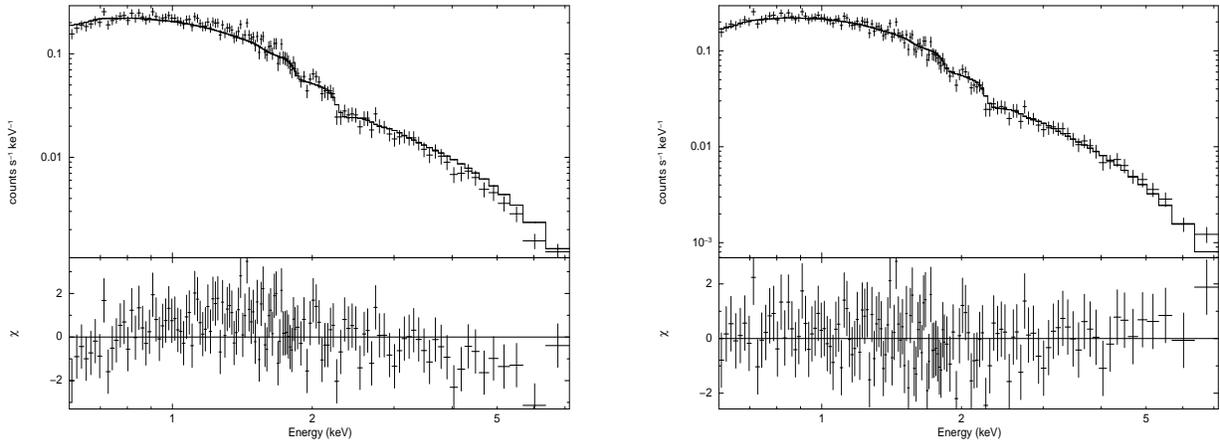

\begin{tabular}{cc}
\includegraphics[height=8.5cm,width=6.5cm,angle=-90]{fig1a.ps}
\label{}
\includegraphics[height=8.5cm,width=6.5cm,angle=-90]{fig1b.ps}
\label{}
\end{tabular}
\caption{An example of the \xmm~spectrum of BZB~J0208+3523 performed on Feb. 14, 2001 (Obs. ID 0084140101)
is here reported to show the goodness of the fitting procedure with the LP model relative to the standard power-law.
\textit{Left}: the systematic deviations on both sides of the residuals from a best fit power-law
with fixed $N_{H,Gal}$ show the need of intrinsic curvature. \textit{Right}: the deviations
disappear on using the LP model with fixed $N_{H,Gal}$.}\label{figura1}
\end{figure*}
%-----------------------------------------------------------------------------------------------------------------------

\subsection{\swf~observations}
The XRT data analysis was performed with the XRTDAS software (v.~2.1), 
developed at the ASI Science Data Center (ASDC) and included in the 
HEAsoft package (v.~6.0.2). 
Event files were calibrated and cleaned with standard filtering 
criteria using the \textsc{xrtpipeline} task. 

Events in the energy range 0.3--10 keV with grades 0--12 (photon counting mode, PC) 
and 0--2 (windowed timing mode, WT) are used in the analyses;
we refer to \citet{hill04} for a description of readout modes, 
and to \citet{burrows05} for a definition of XRT event grades.
This slightly broader band than for \xmm~has no effect on the spectral fits (see \citeauthor{massaro08a}).
For the WT mode data, events were selected for temporal and spectral analysis using a 40 pixel wide 
(1 pixel $=2.36$ $"$) rectangular region centered on the source, and aligned along the 
WT one dimensional stream in sky coordinates. 
Background events were extracted from a nearby source-free rectangular 
region of 40 x 20 pixels. 

For PC mode data, when the source count rate is above 0.45 counts s$^{-1}$, the
data are significantly affected by pile-up in the inner part of the point spread function \citep{moretti05}. 
To remove the pile-up contamination, we extract only events contained in an annular region centered on the source (e.g., \citealt{perri07}). 
The inner radius of the region was determined by comparing the observed  profiles
with the analytical model derived by \citet{moretti05}, and typically has a 4 or 5 pixels radius, 
while the outer radius is 20 pixels for each observation.

For \swf~observations in which the source count rate was below the pile-up limit, 
events are instead extracted using a 20 pixel radius circle. 
The background for PC mode is estimated from a nearby source-free circular region of 20 pixel radius. 

As for \xmm, source spectra are binned 
to ensure a minimum of 30 counts per bin in order to ensure the validity of $\chi^2$ statistics.

\section{X-ray Spectral analysis}
We performed our spectral analysis primarily with the 
{\it Sherpa} \footnote{\underline{http://cxc.harvard.edu/sherpa/}} modeling and fitting application \citep{freeman01}
and we used the {\sc xspec} software package, version 11.3.2 \citep{arnaud96}
as a check of our results.

We describe the X-ray continuum with different spectral models:
1) an absorbed power-law with column density either free, or fixed at the Galactic value $N_{H,Gal}$; 
2) an LP model; 
3) a power-law with an exponential cutoff (PEC) adopting the new expression described below. 
In all models with fixed Galactic column density, we use $N_{H,Gal}$ values 
from the LAB survey \citep{kalberla05} reported in Table 2. 

The LP model in the form:
\begin{equation}\label{eq1}
F(E) = K~E^{-a-b~log(E)}~,
\end{equation}
and the equivalent SED representation used by \citet{tramacere07} and \citeauthor{massaro08a} expressed as:
\begin{equation}\label{eq2}
F(E) = \frac{S_p}{E^2}~{\left(\frac{E}{E_p}\right)}^{-b~log(E/E_{p})}~,
\end{equation}
with $S_p=E_{p}^{2}\, F(E_p)$. 
Both these representations are in units of $photons~cm^{-2}~s^{-1}~keV^{-1}$.
In particular, on using Equation \ref{eq2}, the values of the parameters $E_p$ (the SED energy peak), 
$S_p$ (the SED peak height at $E_p$), and $b$ (the curvature parameter) can be evaluated independently 
in the fitting procedure \citep{massaro06,tramacere07}.

We used the following expression to define the PEC model:
\begin{equation}\label{eq3}
F(E) = \frac{\Sigma_p}{\epsilon_p^2} \left( \frac{E}{\epsilon_p} \right)^{\alpha} exp\left[ \left(1- \frac{E}{\epsilon_p}\right)(2-\alpha) \right]~~~.
\end{equation}
With Equation \ref{eq3}, the three parameters: $\epsilon_p$(the SED energy peak), 
$\Sigma_p$ (the SED height at the peak energy) and the photon index, $\alpha$, can be evaluated independently in the fitting procedure.
We emphasize that the independent estimates of spectral parameters in both LP and PEC models
performed by Equation \ref{eq2} and Equation \ref{eq3}, allow us
to investigate possible correlations among those parameters without the introduction of functional biases.

The results of the LP fits are reported in Appendix; 
the statistical uncertainties quoted refer to the 68\% confidence level (one Gaussian standard deviation).

In some cases, a combination of poor statistics (due to short observational exposures or low count rate),
restricted instrumental energy range, or the location of $E_p$
outside the observational energy range, make it difficult to evaluate the 
spectral curvature. In all these cases the single power-law model is an acceptable description of  the X-ray spectra.
 
For 31 out of the remaining 107 (29\%) of the complete sample of X-ray observations the spectral curvature is consistent with zero
within 1$\sigma$. For 28 out of 135 observations the number of counts did not allow us to perform a good spectral analysis.
In these 59 observations, we added together several low S/N observations for each sources (see the Appendix), and found that 
the co-added spectra are significantly curved in all cases.

\section{Results}
\subsection{X-ray spectral properties}
We present below the results of our X-ray spectral analysis performed on the UBL sample, and compare them with
the known X-ray spectral behavior of TBLs (see \citeauthor{massaro08a}).

We excluded the case of PKS 2155-204 from the TBL sample, because on several 
occasions this source has shown a high energy component dominating over the low energy one (e.g., \citet{aharonian09,abdo11,acciari11}), 
making PKS 2155-204 more similar to a flat spectrum radio quasar than to a HBL.

We also excluded Mrk~421, because it has at least ten times the number of
X-ray observations than any other TBL, and so could dominate the parameter distributions. 

Finally, we excluded from our analysis the giant flare of Mrk 501 in 1997 \citep{massaro06} 
and that of 1H 1426+428 (\citeauthor{massaro08a}),
because we are interested in investigating the spectral behavior in long-term quiescent states, 
rather than in rare, giant, flaring episodes.

We then compared all the UBLs and TBLs observations to search for possible differences 
in their X-ray spectral behavior that could lead to a possible criterion to identify TBL candidates.

Our results are summarized as follows:

1. {\it Spectral models.} We find that
the absorbed power-law model gave unacceptable values of $\chi^2_r$ (i.e., $\chi^2_r \geq 1.5$)
in all cases with sufficient statistics, for which the spectral curvature $b$ could be estimated, 
even when the intrinsic low energy absorption is left as a free parameter.
This model is also inadequate to describe the high energy tail of the X-ray spectra 
above $\sim$ 4 keV (see Figure \ref{figura1} left panel).

Such a lack of intrinsic absorption agrees with the X-ray spectral analyses of TBLs, that are featureless over a broad energy range
(i.e., 0.1 - 10 keV, \citealt{giommi05,perri07,tramacere07}; \citeauthor{massaro08a}). 
An absence of spectral features related to any absorbing material was confirmed by \citet{blustin04}, 
based on the \xmm~RGS spectra.
%-----------------------------------------------------------------------------------------------------------------------
\begin{figure}
\includegraphics[scale=0.32,angle=-90]{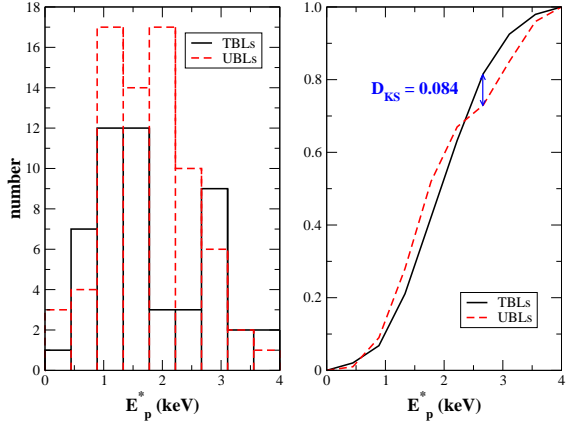}
\caption{The X-ray $E_p$ distribution of UBLs (red) and TBLs (black). 
The sample of TBLs considered here does not include Mrk 421 and PKS 2155-304 
and giant flares of Mrk 501 and 1H 1426+421, as described in Section 5.
The maximum separation $D_{KS}$, of the two cumulative distributions, corresponding to the variable of the KS test, 
is also shown on the plot.}\label{figura2}
\end{figure}
%-----------------------------------------------------------------------------------------------------------------------

On the other hand, both the LP and the PEC models provide acceptable $\chi^2_r$ values for all 
the UBLs (Appendix and Figure 1, right panel), and neither model can be favored over the other
in terms of $\chi^2_r$ and residuals.
We performed a Kolmogorov-Smirnoff (KS) test of the two distributions of $\chi^2_r$ and found
that they are similar at the 99\% level of confidence.

However, it is noteworthy that the $E_p$ values derived using the PEC model
have larger uncertainties than those derived with the LP model. 
This is because with the PEC model, $E_p$ is directly related to the 
exponential cut-off, which is determined by the high energy tail of the 
X-ray spectra, which is not well sampled. 

On the other hand, the LP model provides a systematically better description than PEC function for the TBL X-ray spectra (\citeauthor{massaro08a}).
Thus to compare the TBL and UBL X-ray spectral properties, we adopted the LP model description.
%-----------------------------------------------------------------------------------------------------------------------
\begin{figure}
\includegraphics[scale=0.32,angle=-90]{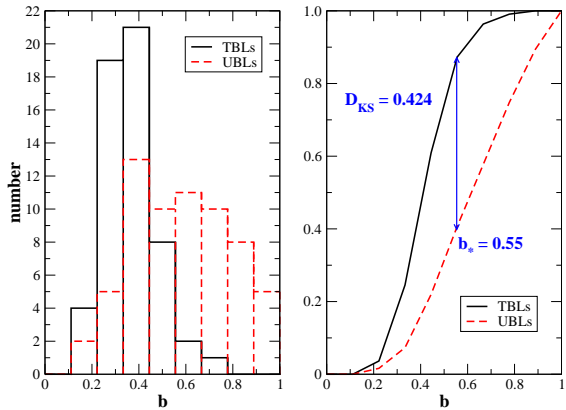}
\caption{The X-ray curvature $b$ distribution of UBLs (red) and TBLs (black). 
The sample of TBLs considered here does not include Mrk 421, PKS 2155-304 
and the giant flares of Mrk 501 and 1H 1426+421, as described in Section 5.
The maximum separation, $D_{KS}$, of the two cumulative distributions (i.e., the variable used for the 
KS test) and the corresponding boundary value of the curvature $b_*$ are 
also shown on the plot.}\label{figura3}
\end{figure}
%-----------------------------------------------------------------------------------------------------------------------

We found the following trends among the spectral parameters:

2. {\it Peak energy $E_p$.} The $E_p$ distribution for the UBLs is consistent with that of TBLs, 
exhibiting a peak around a value $\sim$ 1.75 keV (Figure \ref{figura2}, left panel).
There is a hint of a difference above the $E_p =$ 2.5 keV;
a KS test (Figure \ref{figura2}, right panel) shows that the two distributions do not
differ at a confidence level of 99\%.

In addition, if we identify X-ray flares of HBLs as states where both $E_p$ and $L_p$ increase above their average values,
then the scarcity of high $E_p$ (i.e., higher than $\sim$ 5 keV) values found in our analysis suggests that TBLs are more variable than UBLs,
because in random observations UBLs always appear in their quiescent state.
%-----------------------------------------------------------------------------------------------------------------------
\begin{figure}[!htp]
\includegraphics[scale=0.32,angle=-90]{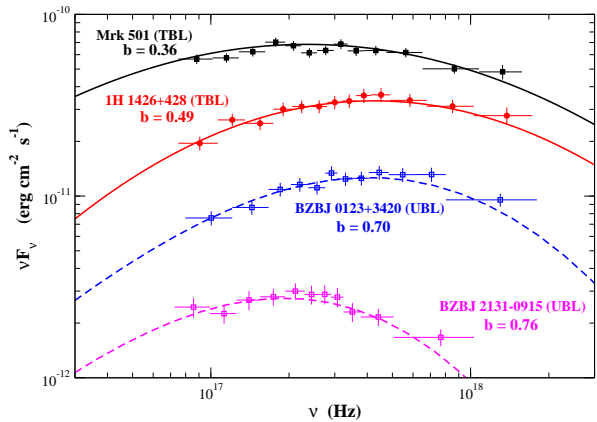}
\caption{The unfolded X-ray SEDs for 4 HBLs: 2 UBLs (dashed lines),  
BZBJ~0123+3420 (blue open squares, 2009-08-28) and BZBJ~2131-0915 (magenta open squares, 2009-03-30),
in comparison with 2 archival observations of the TBLs (solid lines): Mrk 501 (black filled circles, 2006-20-07) 
and 1H 1426+428 (red filled circles, 2006-03-07) (see \citeauthor{massaro08a} for more details).
The TBL X-ray spectra are broader than the UBLs.}\label{figura4}
\end{figure}
%-----------------------------------------------------------------------------------------------------------------------

3. {\it Spectral curvature $b$.} There is a systematic \textit{difference} in $b$ values between TBLs and UBLs (Figure \ref{figura3}, left panel).
It is clear that the curvature in the latter is systematically higher,
indicating that the UBL X-ray spectra spectra are narrower around $E_p$ than those of TBLs.
Applying  a KS test, the two distributions are different at a confidence level of 90\%,
and the maximum separation of the two cumulative distributions of $b$occurs at the boundary value $b_*$ = 0.55 (Figure \ref{figura3}, right panel).
This implies that, given the two $b$ distributions , there is a low probability ($\sim$ 12\%) of finding
to find a TBL with X-ray spectral curvature higher than the boundary value $b_*$ (Figure \ref{figura3}, right panel). 
Thus $b_*$ permit us us to distinguish between TBLs and UBLs based on the X-ray spectral behavior.
The stronger curvature in UBLs is also seen in the acceptable $\chi^2_r$ values when the PEC model is adopted.
This occurs because the PEC model mimics high values of the spectral curvature due 
to its exponential cut-off than a typical LP model with $b \sim$ 0.5.

4. {\it Spectral parameter trends.} There is no clear correlation for the UBLs in the acceleration plane ($E_p$ vs $b$), 
while for TBLs $E_p$ and $b$ anti-correlate (\citeauthor{massaro08a}).
On the other hand, there is no significant trend between $L_p$ and $b$ in either the UBLs or the TBLs (\citeauthor{massaro08a}).
All correlation coefficients evaluated between spectral parameters are lower than 0.1 for both LP and PEC models.

\subsection{Variability}
\begin{figure}[!htp]
\includegraphics[scale=0.32,angle=-90]{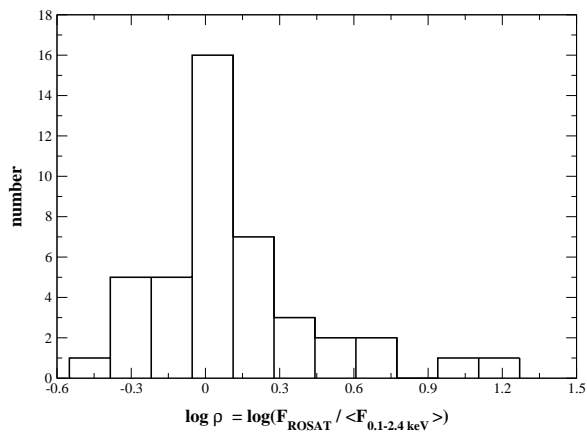}
\caption{The ratio $\rho$ between the ROSAT X-ray flux 
and the one derived from our analysis, for the selected UBLs. 
evaluated in the same energy range (i.e. $<$F$_{0.1-2.4 keV}$$>$)
} 
\label{figura5}
\end{figure}

The UBL X-ray fluxes derived from our archival \swf~and\xmm~ analysis (from December 2004 to October 2010) are consistent 
within a factior of $\sim$2 with those measured, in the same energy range (i.e. 0.1-2.4 keV),  
$\sim$ 15 years earlier ROSAT observations (from June 1990 to February 1999),
as listed in the ROMA BZCAT (\citealt{massaro10b}).
The ROSAT fluxes and those derived from our spectral analysis are reported in Appendix.
Only 18\% of the selected UBLs show a flux ratio: $\rho=<F_{0.1-2.4keV}>/F_{ROSAT}$ higher than 2
(see Figure \ref{figura5}). 
This suggests that UBLs vary little on a 20 year timescale, unlike TBLs which can show variability 
by a factor of $\sim$ 5-10 over 1 year timescale.
%-----------------------------------------------------------------------------------------------------------------------
%-----------------------------------------------------------------------------------------------------------------------

\subsection{\fer~LAT Properties}
The majority, 80\%,of TBLs known up to October 2010 (19 out of 24)
have been also detected in the ``GeV" \fer~ LAT energy range (30 MeV - 100 GeV) \citep{abdo10}.
We searched the \fer~catalog for detections of UBLs and we found that only $\sim$ 20\% (24 out of 118) were detected.
However, for the selected sample of 47 sources investigated here $\sim$ 40\% (19 out of 47) were detected by the \fer~ LAT.
Because the majority of TBLs have been detected by \fer, this could appear to be a requirement for being a TeV source.
However spectral variability may make them undetectable if they lie close to the \fer~detection threshold.

\begin{figure}[!htp]
\includegraphics[scale=0.32,angle=-90]{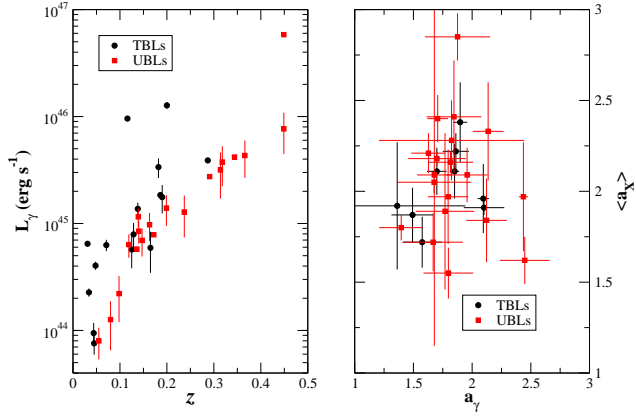}
\caption{The $\gamma$-ray (30 MeV - 100 GeV) luminosity for UBLs (red squares) and TBLs (black circles) 
evaluated using the $\gamma$-ray fluxes reported in the first year \fer~catalog \citep{abdo10}.
The \fer~LAT$\gamma$-ray photon index $\Gamma$ for UBLs (red square)
and TBLs (black circles) with respect to the mean X-ray photon index $<a>$.}\label{figura6}
\end{figure}

We compared the properties of TBLs and UBLs detected by \fer~to see if there are differences
in their $\gamma$-ray properties.
The \fer~LAT ``GeV" luminosity $L_{\gamma}$ vs. redshift is shown in Figure~\ref{figura6}a.
There is a marginal indication that for the \fer~detections 
the UBLs are less luminous than TBLs, in particular at low redshifts, in agreement
with the fact that a most ($\sim$ 50\%) of them in our have not been detected.

The range of values of the $\gamma$-ray spectral index $a_{\gamma}$ is a similar between the TBLs and the UBLs
detected by the \fer~LAT (Figure \ref{figura6}b), the variance of the two distributions are 0.06 and 0.07, respectively. 
Figure \ref{figura6}b shows the $\gamma$-ray photon index $a_{\gamma}$ vs. the average X-ray photon index $<a_X>$ from 
the LP model, weighted with the inverse of the variance.
%-----------------------------------------------------------------------------------------------------------------------
%-----------------------------------------------------------------------------------------------------------------------
We conclude that the MeV-GeV $\gamma$-ray spectral behavior
of the UBLs is similar to that of the TBLs, and the only differences 
appear to reside in the normalization of their $\gamma$-ray flux.
However, this conclusions are valid for those UBLs bright enough in the LAT energy range to be detected by \fer~during one year.
The non-detected HBL could have a different $\gamma$-ray spectral behavior that cannot be investigated with the present data set.

\section{HBLs Detectable at TeV Energies}
From comparing the distribution of the X-ray spectral curvature and the GeV \fer~LAT detections,
we propose criteria to predict which UBLs are more likely to be detectable at TeV energies.

TeV energies lie beyond the inverse Compton peak of the HBL SEDs.
Hence to be detectable they need both a high GeV flux level and a small GeV - TeV spectral curvature.
In the SSC scenario, the X-ray spectral curvature, $b$, of HBLs, evaluated at the synchrotron SED peak, $E_p$,
is a good predictor of the curvature of the inverse Compton peak at GeV - TeV 
energies, although they are not always identical \citep{massaro06}.

We can define three levels of confidence (i.e., TeV classes) in the prediction of TeV detectability (see Table \ref{tabella2}, Col. 12):

Class 1: the best candidates for the future TeV detections are provided by UBLs with 
a GeV \fer~LAT detection and  a curvature, $b$, lower than $b_*$ in all the X-ray observations (see Figure \ref{figura3}b).
We found that four UBLs satisfy both conditions and so are the most likely new TeV detectable extragalactic sources:
BZB J0326+0225, BZB J0442-0018, BZB J0744+7433 and BZB J1743+1953.
Spectral variability  could limit this prediction
but UBLs appear to be less variable in the X-ray band than TBLs (see Section 6.2 and Figure 5).

Class 2: six more UBLs have some X-ray observations with $b < b_*$, and are also detected by \fer~LAT
and so are still TeV candidates:
BZB J0208+3523, BZB J1136+6737, BZB J1417+2543, BZB J1442+1200, BZB J1728+5013, BZB J2250+3824
The variability of $b$ leads us to expect the discovery of other new TBLs when their
X-ray spectrum has $b$ $\leq$ $b_*$.

Class 3: UBLs with $b$ $\leq$ $b_*$ in at least one X-ray observations and $F_X$ $\geq$ 10$^{-11}$ erg s$^{-1}$ cm$^{-2}$ in the 0.5-10 keV energy range, 
but no LAT detection, make up our third class.
The lower GeV normalization makes these less likely TeV candidates.
However, in the single zone SSC scenario \citep[e.g.,][]{paggi09},
the X-ray flux is similar to the detection threshold of 1yr \fer~LAT $\gamma$-ray flux (\citep{atwood09})
and the curvature is as broad as that of TBLs, we suggest that such UBLs can be detected at TeV energies.
Five more UBLs fit class 3:
BZB~J0013-1854, BZB~J0123+3420, BZB~J0214+5144, BZB~J1010-3119 and BZB~J1253-3931.

Our source selection was concluded at the beginning of August 2010. Since then, of the 15 total candidates,
the sources BZB~J1442+1200 and BZB~J2250+3824 from our class 2 and BZB~J0013-1854 and BZB~J1010-3119 
from class 3 have been detected at TeV energies
(see the {\it TeV\,CAT}\footnote{\underline{http://tevcat.uchicago.edu/}} for new announced TeV sources).

%{\bf \citet{costamante02} provided a list of possible TeV candidates on the basis of 7 samples for 
%which only two are included in our selection, namely the Einstein Slew Survey Sample of BL Lacertae Objects and the 
%ROSAT All-Sky Survey-Green Bank BL Lac catalog.
%Their predicted TeV source list was mainly based on the $E_p$ position and
%on the X-ray flux. The TeV candidates listed above have also been defined 
%on the basis of their X-ray spectral shape (i.e., the curvature parameter $b$) and independently
%by optical properties.
%There is an overlap of 8 sources between the list of TeV candidates proposed here and the one of \citet{costamante02}:
%BZB~J0123+3420, BZB~J0214+5144, BZB J0326+0225, BZB J1136+6737, BZB J1417+2543, BZB J1442+1200, BZB J1728+5013, BZB J1743+1953.}

\section{Summary}
We have carried out an extensive X-ray spectral analysis of HBLs to compare 
the spectral behavior of those undetected at TeV energies (UBLs) 
with those already known as TeV emitters (TBLs).
We analyzed all 135 X-ray observations of a sample of 47 UBLs 
present in the \xmm~and \swf~archives up to August 2010.

We found that the $E_p$ distributions of UBLs and TBLs are similar, 
and symmetric around a value of a few keV for both subclasses.
Instead the X-ray spectral curvature, $b$, of UBLs, 
is systematically lower than in TBLs, 
implying that the UBL X-ray spectra are {\textit narrower}.

In addition, in the first year \fer~catalog \citep{abdo10},
we found that the UBL and TBL MeV-GeV $\gamma$-ray spectral behavior is similar, yet
only $\sim$ 40\% of our selected UBLs have been detected in the \fer~LAT energy range vs 80\% of TBLs \citep{abdo10}.

On the basis of our analysis, we have developed criteria to predict likely TBLs.
We present three lists with different levels of confidence for TeV detectability
based on MeV-GeV flux level and keV spectral curvature, comprising a total of 15 TeV candidates. 
By December 2010, four of our candidates have already been detected at TeV energies,
landing support to our selection criteria.

A crucial check for our TeV candidate criteria 
will be provided by X-ray monitoring of candidates from the different TeV classes, 
with simultaneous GeV and TeV observations, 
to investigate the variability timescales of the spectral curvature.

A theoretical interpretation of the $E_p$ and $b$ distributions, for both UBLs and TBLs, 
in terms of systematic and stochastic acceleration mechanisms will be presented in a forthcoming 
paper \citep{massaro11b}.

\acknowledgements
We thank the anonymous referee for useful comments that led to improvements in the paper.
We are grateful to A. Beardmore for discussions regarding the \swf~calibration.
F. Massaro thanks D. Harris for helpful suggestions that improve the presentation and
A. Siemiginowska and B. Refsdal for their useful advices on the use of the Sherpa software package.
The work at SAO was supported by the NASA grant NNX10AD50G.\\
F. Massaro acknowledges the Foundation BLANCEFLOR Boncompagni-Ludovisi, n\'ee Bildt 
for the grant awarded him in 2010 to support his research.
Part of this work is based on archival data, software or on-line services provided by the ASI Science Data Center (ASDC).
This research has made use of data obtained through the High Energy Astrophysics 
Science Archive Research Center Online Service, provided by the NASA/Goddard Space Flight Center.
{\it Facilities:} \facility{\xmm}, \facility{\swf}, \facility{\fer}

{\it Note added in proof.} The source BZBJ1743+1935 (i.e.,
1ES 1741+196) indicated, on the basis of our investigation, as a
TeV candidate of class I, has been recently discovered at TeV energies
as predicted by our study (see the {\it TeV\,CAT}\footnote{\underline{http://tevcat.uchicago.edu/}}
for more details). This observation supports our selection criteria
for TeV candidates in the HBL subclass.

~
{}

\appendix 
\section{Results of the X-ray spectral analysis for the UBLs}
The following tables report the log of the selected X-ray observations and the values of the spectral parameters 
we have derived for UBLs in our sample. 

In \swf~Tables the column $Frame$ reports on the observation modality 
(PC for photon counting and WT for windowed timed, see also Section 3.2 for details), 
and $Exps$ means the exposure time in seconds.

In \xmm~Table, $Frame$ indicates the EPIC camera used (M1$=$MOS1 and M2$=$MOS2), 
the modes (PW$=$partial window and FW$=$full window) and the filter (Th$=$thin, Md$=$medium, Tk$=$thick) 
used for each pointing (see Section 2.2 for details), and the exposure is reported in seconds in the column $Exps$. 

All other columns in each table refer to bestfit with the log-parabolic model. 
When the value estimated for a spectral parameter is consistent with zero in a $2\sigma$ interval, 
the values reported in each table refer to the power-law model bestfit (see Section 4). 
In these cases, the curvature parameter $b$, the SED peak energy $E_p$ and 
the corresponding SED peak height $S_p$ cannot be reliably evaluated, and are marked with a dashed line.

Values of $E_p$ are reported in keV, the normalization $K$ in units of 10$^{-4}$~photons~cm$^{-2}$~s$^{-1}$~keV$^{-1}$
 and $S_p$ in units of  10$^{-13}$ erg~cm$^{-2}$~s$^{-1}$ 
with $F_X$ denoting the 0.5-10 keV flux measured in units of  10$^{-11}$ erg~cm$^{-2}$~s$^{-1}$.
 
\begin{table*}
\caption{\textit{Swift} ~spectral analysis results with the LP model of the UBLs.}
\tiny
\begin{tabular}{|llllccccccc|}
\hline
\noalign{\smallskip}
Obs ID & $Date$ & Frame & Exps & $a$ & $b$ & $E_p$ & $K$ & $S_p$ & $F_X$ & $\chi^2_{r}$\\ % flux swift 0.1-2.4 e ratio Frosat/Fswift nella stessa banda    <a_X> e <Gamma_G>
\hline
\noalign{\smallskip}
\hline
\textbf{BZB~J0013-1854} &          &    &       &            &            &            &       &            &      &          \\
00031806002             & 10/09/10 & pc &  3782 & 1.83(0.06) & 0.68(0.15) & 1.34(0.11) & 34(1) &  52.2(2.1) & 1.24 & 1.43(36) \\ % ok 9.31e-12
00031806003             & 10/09/10 & pc &  4193 & 1.89(0.05) & 0.46(0.13) & 1.31(0.14) & 30(1) &  48.1(1.7) & 1.16 & 1.23(41) \\ % ok 8.57e-12 
sum                     & -        & pc &  7975 & 1.86(0.04) & 0.59(0.09) & 1.31(0.07) & 33(1) &  53.8(1.8) & 1.24 & 1.31(75) \\ % ok 9.29e-12
\noalign{\smallskip}
00031806004             & 10/09/12 & pc &  3848 & 1.88(0.05) & 0.64(0.15) & 1.24(0.11) & 32(1) &  52.7(2.1) & 1.18 & 1.05(33) \\ % ok 9.10e-12 
00031806005             & 10/09/18 & pc &  2744 & 1.88(0.06) & 0.36(0.19) & 1.44(0.29) & 28(1) &  45.0(2.1) & 1.14 & 1.20(24) \\ % ok 8.09e-12     6.488e-12/8.7675e-12=0.74
sum                     & -        & pc &  6592 & 1.87(0.04) & 0.59(0.09) & 1.28(0.08) & 30(1) &  49.3(1.4) & 1.13 & 0.67(59) \\ % ok 8.54e-12
\hline
\textbf{BZB~J0123+3420} &          &    &       &            &            &            &       &            &      &          \\
00035000001	            & 05/06/09 & pc &  1587 & 1.62(0.10) & 0.64(0.21) & 1.99(0.31) & 85(5) & 155.0(9.1) & 3.56 & 0.70(15) \\ % ok 20.5e-12
00035000002	            & 06/07/06 & pc &  5414 & 1.70(0.06) & 0.46(0.12) & 2.11(0.28) & 65(2) & 115.8(4.1) & 2.80 & 1.02(43) \\ % ok 15.8e-12
00035000003	            & 06/07/10 & pc &  2444 & 1.74(0.11) & 0.33(0.22) & 2.46(0.89) & 70(4) & 126.5(7.2) & 3.18 & 1.61(18) \\ % ok 17.5e-12
sum                     & -        & pc &  9445 & 1.63(0.04) & 0.57(0.09) & 2.13(0.17) & 66(2) & 122.5(3.2) & 2.88 & 0.98(77) \\ % ok 16.2e-12
\noalign{\smallskip}
00030876001	            & 07/01/18 & pc &   782 & 1.39(0.20) & 1.28(0.38) & 1.73(0.18) & 99(7) & 188.5(14.8)& 3.63 & 0.82(7)  \\ % ok 22.9e-12
\noalign{\smallskip}
00035000005	            & 07/09/07 & pc &  1018 & 1.69(0.19) & -          &          - & 72(6) & -          & 3.30 & 0.70(6)  \\ % ok 17.8e-12
00035000006	            & 07/09/11 & pc &   521 & -          & -          & -          & -     & -          & -    &          \\ % ok
00035000007	            & 07/09/12 & pc &   521 & -          & -          & -          & -     & -          & -    &          \\ % ok
00035000008	            & 07/09/13 & pc &   511 & -          & -          & -          & -     & -          & -    &          \\ % ok
00035000004             & 07/09/14 & pc &   699 & -          & -          & -          & -     & -          & -    &          \\ % ok
sum                     & -        & pc &  3271 & 1.69(0.06) & 0.76(0.14) & 1.59(0.11) & 76(3) & 130.2(4.7) & 2.81 & 1.09(39) \\ % ok 17.9e-12
\noalign{\smallskip}
00035000009	            & 07/09/15 & pc &  1289 & 1.60(0.18) & 0.88(0.51) & 1.57(0.24) & 80(6) & 143.4(13.5)& 3.00 & 0.44(7)  \\ % ok 18.7e-12
00035000010             & 07/09/27 & pc &   754 & 1.69(0.24) & 0.93(0.65) & 1.46(0.24) &166(14)& 282.1(26.1)& 5.71 & 1.08(5)  \\ % ok 38.5e-12
00035000011	            & 07/10/27 & pc &  1479 & 1.94(0.14) & -          & -          & 65(4) & -          & 3.00 & 1.11(9)  \\ % ok 17.0e-12
sum                     & -        & pc &  3521 & 1.74(0.07) & 0.70(0.17) & 1.52(0.13) & 77(3) & 129.9(5.5) & 2.83 & 1.43(30) \\ % ok 18.3e-12
\noalign{\smallskip}
00035000013	            & 07/11/09 & pc &  2105 & 1.80(0.10) & 0.53(0.22) & 1.53(0.23) & 90(5) & 150.0(8.4) & 3.45 & 0.96(17) \\ % ok 11.7e-12
00035000014	            & 07/11/16 & pc &  1380 & 1.67(0.15) & 0.52(0.37) & 2.07(0.70) & 84(6) & 152.3(10.6)& 3.62 & 1.80(10) \\ % ok 20.6e-12
sum                     & -        & pc &  3485 & 1.78(0.06) & 0.52(0.13) & 1.62(0.16) & 89(3) & 149.8(5.4) & 3.48 & 0.95(40) \\ % ok 21.5e-12
\noalign{\smallskip}
00037298001	            & 08/03/06 & pc &  1248 & 1.72(0.11) & 0.33(0.24) & 2.69(1.34) & 73(4) & 134.2(8.2) & 3.05 & 0.83(17) \\ % ok 19.9e-12
00037298002	            & 08/06/08 & pc &  4757 & 1.69(0.06) & 0.55(0.13) & 1.92(0.20) & 69(2) & 122.6(4.4) & 2.88 & 1.56(43) \\ % ok 16.9e-12
00037298003	            & 09/08/28 & pc &  4816 & 1.68(0.06) & 0.70(0.13) & 1.70(0.13) & 74(2) & 129.6(4.4) & 2.88 & 0.57(45) \\ % ok 17.7e-12     25.254e-12/19.6538e-12=1.28
sum                     & -        & pc &       & 1.69(0.04) & 0.63(0.07) & 1.78(0.09) & 75(2) & 131.4(3.0) & 2.99 & 0.99(96) \\ % ok 18.0e-12
\hline
\textbf{BZB~J0201+0034} &          &    &       &            &            &            &       &            &      &          \\
00038117001	            & 09/06/05 & pc &  5041 & 1.94(0.08) & 0.69(0.28) & 1.10(0.14) & 12(1) &  18.8(1.1) & 0.40 & 1.34(16) \\ % ok 3.26e-12     3.526e-12/3.26e-12=1.08
\hline
\textbf{BZB~J0214+5144} &          &    &       &            &            &            &       &            &      &          \\
00038333001	            & 08/12/10 & pc &  5040 & 1.72(0.07) & 0.65(0.13) & 1.62(0.12) & 49(1) &  83.6(2.6) & 1.67 & 0.96(53) \\ % ok 8.87e-12
00038333002             & 08/12/11 & pc &   780 & -          & -          & -          & -     & -          & -    &          \\ % ok
00038333003	            & 09/09/16 & pc &  3376 & 1.65(0.09) & 0.43(0.15) & 2.58(0.45) & 47(2) &  89.2(3.4) & 1.99 & 0.99(40) \\ % ok 8.93e-12     4.579e-12/8.449e-12=0.54
sum                     & -        & pc &  9196 & 1.66(0.05) & 0.61(0.09) & 1.90(0.11) & 47(1) &  83.3(1.9) & 1.73 & 1.10(95) \\ % ok 8.63e-12
\hline
\textbf{BZB~J0227+0202} &          &    &       &            &            &            &       &            &      &          \\
00037512001	            & 08/06/24 & pc &  1308 & -          & -          & -          & -     & -          & -    &          \\ % ok
00037512002             & 10/08/19 & pc &  2847 & 1.65(0.09) & 0.70(0.23) & 1.79(0.26) & 26(1) &  46.9(2.6) & 1.09 & 0.66(17) \\ % ok 7.03e-12     18.184e-12/7.03e-12=2.59
sum                     & -        & pc &  4156 & 1.73(0.08) & 0.49(0.23) & 1.89(0.45) & 16(1) &  27.1(1.4) & 0.67 & 0.66(19) \\ % ok 4.28e-12
\hline 
\textbf{BZB~J0325-1646} &          &    &       &            &            &            &       &            &      &          \\
00035005001             & 05/06/29 & pc &  7671 & 2.85(0.13) & -          & -          &3.9(0.3)& -         & 0.11 & 1.26(8)  \\ % ok 1.48e-12
00035005002             & 06/07/19 & pc &  1951 & -          & -          & -          & -     & -          & -    &          \\ % ok
00035005003             & 07/06/08 & pc &   346 & -          & -          & -          & -     & -          & -    &          \\ % ok              27.164e-12/1.48e-12=18.35      2.85
sum                     & -        & pc &  9969 & 2.92(0.11) & -          & -          &3.7(0.2)& -         & 0.10 & 1.07(11) \\ % ok 1.43e-12
\hline
\textbf{BZB~J0326+0225} &          &    &       &            &            &            &       &            &      &          \\
00035006001             & 05/06/26 & pc & 10680 & 2.28(0.08) & 0.42(0.28) & 0.52(0.21) &7.5(0.4)& 13.5(1.5) & 0.21 & 0.64(19) \\ % ok 1.64e-12
00035006002             & 05/06/29 & pc &  4650 & 2.40(0.20) & -          & -          & 7(1)  & -          & 0.18 & 0.42(5)  \\ % ok 1.45e-12
00035006003             & 05/07/11 & pc &  6549 & 2.31(0.16) & -          & -          &4.7(0.4)& -         & 0.16 & 1.04(5)  \\ % ok 1.09e-12     12.038e-12/1.36e-12=8.85       2.33
sum                     & -        & pc & 21880 & 2.26(0.06) & 0.43(0.14) & 0.49(0.16) &6.4(0.2)& 11.3(0.7) & 0.18 & 0.86(35) \\ % ok 1.40e-12                 
\hline
\textbf{BZB~J0441+1504} &          &    &       &            &            &            &       &            &      &          \\
00036806002	            & 08/01/08 & pc &  7059 & 1.18(0.15) & 1.17(0.26) & 2.23(0.17) & 18(1) &  40.4(1.7) & 0.76 & 1.26(31) \\ % ok 3.62e-12     10.195e-12/3.62e-12=2.82
\hline
\textbf{BZB~J0442-0018} &          &    &       &            &            &            &       &            &      &          \\
00036312001             & 07/03/28 & pc &  1275 & -          & -          & -          & -     & -          & -    &          \\ % ok
00036312003             & 07/04/17 & pc &  5259 & -          & -          & -          & -     & -          & -    &          \\ % ok
00036312003             & 07/07/30 & pc &  2776 & -          & -          & -          & -     & -          & -    &          \\ % ok
00036312005             & 08/10/19 & pc & 10780 & 1.97(0.30) & -          & -          &2.1(0.2)& -         & 0.09 & 0.28(3)  \\ % ok 0.55e-12     1.673e-12/0.55e-12=3.04       1.97
sum                     & -        & pc & 20090 & 1.87(0.13) & 0.45(0.32) & 1.39(0.33) &2.3(0.1)&  3.8(0.3) & 0.09 & 0.84(10) \\ % ok 0.57e-12
\noalign{\smallskip}
\hline
\end{tabular}\\
Col. (7) $E_p$ is in keV. \\
Col. (8) $K$ is in 10$^{-4}$~photons~cm$^{-2}$~s$^{-1}$~keV$^{-1}$.\\
Col. (9) $S_p$ is in units of 10$^{-13}$ erg~cm$^{-2}$~s$^{-1}$.\\
Col. (10) $F_X$ denoting the 0.5-10 keV flux measured in units of 10$^{-11}$ erg~cm$^{-2}$~s$^{-1}$. 
\end{table*}

\begin{table*}
\caption{\textit{Swift} ~spectral analysis results with the LP model of the UBLs.}
\tiny
\begin{tabular}{|llllccccccc|}
\hline
\noalign{\smallskip}
Obs ID & $Date$ & Frame & Exps & $a$ & $b$ & $E_p$ & $K$ & $S_p$ & $F_X$ & $\chi^2_{r}$\\ % flux swift 0.1-2.4 e ratio Frosat/Fswift nella stessa banda    <a_X> e <Gamma_G>
\hline
\noalign{\smallskip}
\hline
\textbf{BZB~J0621-3411} &          &    &       &            &            &            &       &            &      &          \\
00038819001             & 09/07/29 & pc &   949 & 0.90(0.52) & 1.47(1.14) & 2.37(0.78) & 17(2) & 43.4(5.1)  & 0.84 & 0.63(2)  \\ % ok 4.34e-12     1.838e-12/4.34e-12=0.43
\hline
\textbf{BZB~J0751+1730} &          &    &       &            &            &            &       &            &      &          \\
00036808001             & 07/05/30 & pc &  3102 & 1.20(0.41) & 1.33(0.88) & 2.01(0.43) &  6(1) &  13.6(1.5) & 0.27 & 0.09(4)  \\ % ok 1.53e-12     1.781e-12/1.53e-12=1.16
\hline
\textbf{BZB~J0753+2921} &          &    &       &            &            &            &       &            &      &          \\
00036809001             & 08/03/06 & pc &  4500 & -          & -          & -          & -     & -          & -    &          \\ % ok
\hline 
\textbf{BZB~J0847+1133} &          &    &       &            &            &            &       &            &      &          \\
00037396001             & 08/02/29 & pc &  2022 & 1.80(0.07) & -          & -          & 33(2) & -          & 1.63 & 0.85(20) \\ % ok 9.39e-12     11.006e-12/9.39e-12=1.17        1.80
\hline
\textbf{BZB~J0916+5238} &          &    &       &            &            &            &       &            &      &          \\
00038165001	            & 09/03/07 & pc &  7710 & 2.08(0.06) & 0.55(0.18) & 0.84(0.14) &9.5(0.4)& 15.4(0.7) & 0.32 & 0.65(23) \\ % ok 2.98e-12     3.833e-12/2.98e-12=1.29
\hline
\textbf{BZB~J0930+4950} &          &    &       &            &            &            &       &            &      &          \\
00039154001             & 10/10/12 & pc &  2869 & 1.67(0.06) & 0.89(0.15) & 1.53(0.11) & 32(1) &  54.8(2.3) & 1.20 & 0.73(29) \\ % ok 8.85e-12     16.672e-12/8.85e-12=1.88
\hline
\textbf{BZB~J0952+7502} &          &    &       &            &            &            &       &            &      &          \\
00036810001	            & 07/05/20 & pc &  9759 & 1.80(0.07) & 0.44(0.16) & 1.71(0.30) &7.3(0.3)& 12.3(0.6) & 0.31 & 0.89(23) \\ % ok 2.09e-12
00036810002             & 07/10/04 & pc &  2230 & -          & -          & -          & -     & -          & -    &          \\ % ok              2.485e-12/2.09e-12=1.18
sum                     & -        & pc & 11990 & 1.79(0.07) & 0.58(0.17) & 1.51(0.17) &7.1(0.3)& 11.9(0.6) & 0.28 & 1.30(26) \\ % ok 1.99e-12
\hline
\textbf{BZB~J1010-3119} &          &    &       &            &            &            &       &            &      &          \\
00030940002	            & 07/05/17 & pc &  1481 & 2.00(0.09) & 0.46(0.23) & 1.00(0.24) & 43(2) &  68.9(3.4) & 1.41 & 0.85(18) \\ % ok 9.14e-12
00030940004	            & 07/05/18 & pc &  1968 & 1.82(0.10) & 1.13(0.25) & 1.21(0.09) & 42(2) &  68.9(3.4) & 1.18 & 1.08(20) \\ % ok 8.45e-12     10.149e-12/8.794e-12=1.15
sum                            & -        & pc &  3700 & 1.88(0.06) & 0.76(0.14) & 1.21(0.09) & 42(1) &  67.4(2.4) & 1.29 & 1.40(39) \\ % ok 8.62e-12
\hline
\textbf{BZB~J1022+5124} &          &    &       &            &            &            &       &            &      &          \\
00036811001	            & 08/01/13 & pc &  5703 & 1.60(0.12) & -          & -          &8.6(0.5)& -         & 0.49 & 1.61(14) \\ % ok 2.75e-12     3.442e-12/2.75e-12=1.25
\hline
\textbf{BZB~J1053+4929} &          &    &       &            &            &            &       &            &      &          \\
00031594001	            & 10/01/21 & pc &  5243 & 2.21(0.11) & 1.01(0.46) & 0.79(0.12) &6.4(0.5)& 10.5(0.8) & 0.18 & 1.67(8)  \\ % ok 1.86e-12     0.821e-12/1.86e-12=0.44         1.65
\hline
\textbf{BZB~J1056+0252} &          &    &       &            &            &            &       &            &      &          \\
00037547001	            & 07/06/08 & pc &  4725 & 1.65(0.07) & 0.59(0.15) & 1.97(0.25) & 20(1) &  35.6(1.5) & 0.84 & 1.02(29) \\ % ok 5.07e-12     6.903e-12/5.07e-12=1.36
\hline
\textbf{BZB~J1111+3452} &          &    &       &            &            &            &       &            &      &          \\
00038219001	            & 09/04/18 & pc &  4590 & -          & -          & -          & -     & -          & -    &          \\ % ok
\hline
\textbf{BZB~J1117+2014} &          &    &       &            &            &            &       &            &      &          \\
00038451001	            & 09/04/20 & pc &  1341 & 2.40(0.13) & 0.69(0.59) & 0.52(0.32) & 22(2) &  39.3(4.8) & 0.61 & 0.72(6) \\ % ok 7.22e-12      33.576e-12/7.22e-12=4.65
\hline
\textbf{BZB~J1136+6737} &          &    &       &            &            &            &       &            &      &          \\
00037135001	            & 07/05/25 & pc &  2788 & 1.60(0.09) & 0.96(0.23) & 1.60(0.14) & 49(3) &  85.8(4.8) & 1.87 & 0.98(17) \\ % ok 13.6e-12
00037135002	            & 07/05/30 & pc &  4487 & 1.74(0.07) & 0.68(0.19) & 1.54(0.16) & 44(2) &  74.3(3.6) & 1.74 & 0.71(22) \\ % ok 13.0e-12
00037135003	            & 08/02/16 & pc &  4291 & 1.49(0.05) & 0.68(0.10) & 2.37(0.23) & 95(3) & 189.2(5.7) & 4.56 & 0.85(62) \\ % ok 27.7e-12
00036812001	            & 08/01/29 & pc &  3983 & 1.47(0.07) & 0.42(0.14) & 4.26(1.57) & 49(2) & 115.4(8.3) & 2.83 & 1.00(28) \\ % ok 14.9e-12
00036812002	            & 08/01/30 & pc &  2304 & 1.66(0.09) & 0.38(0.19) & 2.40(0.72) & 51(3) &  89.1(4.8) & 2.56 & 1.06(16) \\ % ok 15.9e-12     14.752e-12/17.02e-12=0.87
\hline
\textbf{BZB~J1145-0340} &          &    &       &            &            &            &       &            &      &          \\
00036813001             & 07/11/09 & pc &  2870 & -          & -          & -          & -     & -          & -    &          \\ % ok
00036813002	            & 07/12/06 & pc &  2934 & 1.64(0.17) & 1.26(0.43) & 1.39(0.15) &  9(1) &  14.6(1.3) & 0.28 & 0.57(5)  \\ % ok 2.16e-12     4.101e-12/2.16e-12=1.90
sum                     & -        & pc &  5805 & 1.87(0.11) & 1.02(0.37) & 1.16(0.13) &  8(1) &  12.5(0.9) & 0.25 & 1.21(10) \\ % ok 2.03e-12
\hline
\textbf{BZB~J1154-0010} &          &    &       &            &            &            &       &            &      &          \\
00038231001             & 09/11/08 & pc &  1209 & -          & -          & -          & -     & -          & -    &          \\ % ok
\hline
\textbf{BZB~J1237+6258} &          &    &       &            &            &            &       &            &      &          \\
00042002001             & 06/11/12 & pc &  3114 & -          & -          & -          & -     & -          & -    &          \\ % ok
00042002008             & 06/02/15 & pc &  6501 & 2.10(0.06) & 0.44(0.22) & 0.77(0.16) & 12(1) &  20.0(0.8) & 0.43 & 1.02(25) \\ % ok 4.18e-12
00060001001             & 06/11/14 & pc &  2677 & -          & -          & -          & -     & -          & -    &          \\ % ok
sum                     & -        & pc & 12290 & 2.17(0.05) & 0.47(0.18) & 0.6590.14) &7.5(0.3)& 12.5(0.48)& 0.25 & 0.99(30) \\ % ok 2.60e-12
\noalign{\smallskip}
00066002001             & 07/04/18 & pc &   748 & -          & -          & -          & -     & -          & -    &          \\ % ok
00042002024             & 07/03/22 & pc &  3664 & 2.10(0.12) & -          & -          &  6(1) & -          & 2.28 & 0.70(5)  \\ % ok 2.17e-12
00042002018             & 07/06/12 & pc &  1593 & 1.80(0.18) & 1.02(0.75) & 1.25(0.26) & 12(1) &  19.4(2.1) & 0.40 & 0.33(3)  \\ % ok 3.37e-12
00042002028             & 07/10/10 & pc &  1124 & -          & -          & -          & -     & -          & -    & -        \\ % ok
sum                     & -        & pc &  7129 & 1.93(0.08) & 0.59(0.20) & 1.15(0.15) &8.2(0.4)& 13.2(0.7) & 0.30 & 0.48(18) \\ % ok 2.58e-12
\noalign{\smallskip}
00042002027             & 08/03/03 & pc &  1845 & -          & -          & -          & -     & -          & -    &          \\ % ok
00038250001             & 09/03/10 & pc &  2769 & -          & -          & -          & -     & -          & -    &          \\ % ok
00038250002             & 09/04/23 & pc &  3062 & 2.54(0.34) & -          & -          &2.7(0.6)& -         & 0.09 & 0.65(3)  \\ % ok 1.28e-12  
00042002034             & 09/10/25 & pc &   910 & -          & -          & -          & -     & -          & -    &          \\ % ok
sum                     & -        & pc &  8587 & 2.26(0.11) & 0.82(0.44) & 0.70(0.16) &3.9(0.3)&  6.6(0.5) & 0.11 & 1.18(9)  \\ % ok 1.28e-12
\noalign{\smallskip}
00042002035             & 10/01/27 & pc &  4729 & 2.07(0.14) & -          & -          &5.1(0.4)& -         & 0.19 & 0.79(5)  \\ % ok 1.75e-12
00042002038             & 10/10/13 & pc &  1568 & -          & -          & -          & -     & -          & -    &          \\ % ok              2.284e-12/2.549e-12=0.90
sum                     & -        & pc &  8587 & 1.98(0.13) & 0.68(0.42) & 1.03(0.22) &4.2(0.4)&  6.8(0.6) & 0.15 & 0.58(7)  \\ % ok 1.32e-12
\noalign{\smallskip}
\hline
\end{tabular}\\
Col. (7) $E_p$ is in keV. \\
Col. (8) $K$ is in 10$^{-4}$~photons~cm$^{-2}$~s$^{-1}$~keV$^{-1}$.\\
Col. (9) $S_p$ is in units of 10$^{-13}$ erg~cm$^{-2}$~s$^{-1}$.\\
Col. (10) $F_X$ denoting the 0.5-10 keV flux measured in units of 10$^{-11}$ erg~cm$^{-2}$~s$^{-1}$. 
\end{table*}

\begin{table*}
\caption{\textit{Swift} ~spectral analysis results with the LP model of the UBLs.}
\tiny
\begin{tabular}{|llllccccccc|}
\hline
\noalign{\smallskip}
Obs ID & $Date$ & Frame & Exps & $a$ & $b$ & $E_p$ & $K$ & $S_p$ & $F_X$ & $\chi^2_{r}$\\ % flux swift 0.1-2.4 e ratio Frosat/Fswift nella stessa banda    <a_X> e <Gamma_G>
\hline
\noalign{\smallskip}
\hline
\textbf{BZB~J1253-3931} &          &    &       &            &            &            &       &            &      &          \\
00037538001	            & 08/12/20 & pc &  4651 & 1.46(0.11) & 0.41(0.21) & 4.68(2.68) & 20(1) &  47.5(4.7) & 1.08 & 0.65(25) \\ % ok 4.59e-12     7.046e-12/4.59e-12=1.54
\hline
\textbf{BZB~J1257+2412} &          &    &       &            &            &            &       &            &      &          \\
00031203001	            & 08/05/09 & pc &  2200 & 1.89(0.10) & 0.55(0.28) & 1.25(0.22) & 17(1) &  28.3(1.9) & 0.66 & 0.98(11) \\ % ok 5.35e-12     7.221e-12/5.35e-12=1.35
\hline
\textbf{BZB~J1341+3959} &          &    &       &            &            &            &       &            &      &          \\
00038268001	            & 08/10/15 & pc &  5198 & 1.63(0.06) & 0.70(0.14) & 1.84(0.18) & 23(1) &  41.2(1.5) & 0.98 & 0.96(37) \\ % ok 6.87e-12
00038268002	            & 09/12/21 & pc &  5901 & 1.64(0.07) & 0.70(0.16) & 1.82(0.19) &14.0(0.6)&25.1(1.2) & 0.60 & 1.42(26) \\ % ok 4.20e-12
sum                     & -        & pc & 11100 & 1.66(0.04) & 0.66(0.10) & 1.82(0.14) &18.1(0.5)&32.3(0.1) & 0.78 & 1.04(63) \\ % ok 5.49e-12
\noalign{\smallskip}
00040599001             & 10/10/10 & pc &   854 & -          & -          & -          & -     & -          & -    & -        \\ % ok  
00040599002             & 10/10/15 & pc &  3947 & 1.55(0.09) & 0.66(0.21) & 2.20(0.38) & 13(1) &  25.7(1.5) & 0.63 & 1.03(17) \\ % ok 4.03e-12     5.152e-12/5.03e-12=1.02
sum                     & -        & pc &  4801 & 1.60(0.08) & 0.50(0.20) & 2.53(0.75) & 13(1) &  25.0(1.4) & 0.64 & 1.42(21) \\ % ok 4.01e-12
\hline
\textbf{BZB~J1417+2543} &          &    &       &            &            &            &       &            &      &          \\
00035270001	            & 05/12/20 & pc &  8547 & 1.83(0.04) & 0.43(0.08) & 1.59(0.14) & 64(2) & 106.6(2.9) & 2.69 & 0.92(69) \\ % ok 19.3e-12
00056620002             & 05/05/26 & pc &   775 & 1.72(0.15) & 0.88(0.59) & 1.44(0.34) & 84(7) & 142.1(11.6)& 3.08 & 0.57(6)  \\ % ok 23.3e-12
00056620002	           & 26/05/05 & wt &  1015 & 1.87(0.07) & 0.51(0.19) & 1.34(0.18) & 81(4) & 132.8(6.2) & 1.52 & 0.75(25)  \\
00035270002	            & 06/07/11 & pc &  1882 & 1.90(0.09) & -          & -          & 60(4) & -          & 2.93 & 0.58(13) \\ % ok 19.8e-12
\noalign{\smallskip}
00031204001	            & 08/05/10 & pc &  1694 & 1.75(0.10) & 0.39(0.34) & 2.08(1.24) & 51(2) &  86.8(6.7) & 2.25 & 1.78(10) \\ % ok 14.9e-12
00031204002	            & 08/05/30 & pc &  1664 & 2.03(0.11) & 0.63(0.38) & 0.94(0.21) & 49(4) &  78.3(5.6) & 1.64 & 0.39(8)  \\ % ok 14.7e-12     15.304e-12/18.40e-12=0.83
sum                     & -        & pc &  3358 & 1.87(0.07) & 0.53(0.19) &            & 49(2) &            & 1.91 & 1.36(23) \\ % ok 14.7e-12
\hline
\textbf{BZB~J1439+3932} &          &    &       &            &            &            &       &            &      &          \\
00037514002             & 08/10/15 & pc &  1458 & 2.49(0.16) & -          & -          & 17(2) & -          & 0.56 & 0.38(5)  \\ % ok 7.31e-12
00037514001             & 08/06/07 & pc &   832 & 2.18(0.27) & -          & -          & 19(3) & -          & 0.78 & 1.86(2)  \\ % ok 7.34e-12     11.076e-12
sum                     & -        & pc &  2290 & 2.42(0.08) & 0.64(0.29) & 0.47(0.17) & 18(1) &  33.6(2.6) & 0.52 & 0.82(12) \\ % ok 6.34e-12
\hline
\textbf{BZB~J1442+1200} &          &    &       &            &            &            &       &            &      &          \\
00031218002	            & 08/06/12 & pc &  1732 & 1.83(0.11) & 0.37(0.34) & 1.70(0.68) & 40(2) &  66.9(4.8) & 1.73 & 1.27(9)  \\ % ok 12.2e-12
00031218003             & 10/02/26 & pc &  1110 & 2.15(0.18) & -          & -          & 42(5) & -          & 1.59 & 0.12(3)  \\ % ok 14.2e-12
00031218005             & 10/03/09 & pc &  1058 & 1.80(0.20) & 1.36(0.57) & 1.18(0.15) & 43(4) &  70.2(6.4) & 1.27 & 1.79(3)  \\ % ok 11.1e-12
00040617002             & 10/12/09 & pc &  3308 & 1.82(0.32) & 0.32(0.16) & 1.93(0.52) & 56(2) &  94.7(4.3) & 2.50 & 0.94(27) \\ % ok 17.1e-12     7.82e-12/7.324e-12=1.07
sum                     & -        & pc &  7208 & 1.85(0.05) & 0.48(0.11) & 1.43(0.13) & 51(2) &  83.3(2.7) & 2.03 & 0.90(49) \\ % ok 15.2e-12
\hline
\textbf{BZB~J1534+3715} &          &    &       &            &            &            &       &            &      &          \\
00038300001             & 08/12/12 & pc & 14760 & 2.87(0.15) & -          & -          &1.3(0.1)& -         & 0.04 & 0.85(4)  \\ % ok 0.69e-12     0.228e-12/0.69e-12=0.33
\hline
\textbf{BZB~J1605+5421} &          &    &       &            &            &            &       &            &      &          \\
00038303001	            & 09/01/18 & pc &  7066 & 1.37(0.12) & 0.84(0.33) & 2.36(0.61) &5.0(0.3)& 10.5(0.8) & 0.24 & 0.94(10) \\ % ok 1.45e-12     3.874e-12/1.45e-12=2.67
\hline
\textbf{BZB~J1728+5013} &          &    &       &            &            &            &       &            &      &          \\
00040635002	            & 10/04/02 & pc &   260 & -          & -          & -          & -     & -          & -    & -        \\ % ok
00040635001	            & 10/04/05 & pc &  1395 & 1.98(0.10) & 0.60(0.34) & 1.03(0.19) & 52(3) &  82.7(5.7) & 1.78 & 0.57(10) \\ % ok 14.5e-12
00040635003	            & 10/04/05 & pc &  2150 & 2.12(0.08) & -          & -          & 45(3) & -          & 1.66 & 0.93(14) \\ % ok 13.9e-12
00040635004	            & 10/05/01 & pc &  1667 & 2.12(0.08) & 0.64(0.26) & 0.81(0.14) & 38(2) &  62.1(3.3) & 1.21 & 1.02(16) \\ % ok 10.9e-12     20.374e-12/13.1e-12=1.55
sum                     & -        & pc &  5473 & 2.12(0.05) & 0.46(0.15) & 0.74(0.15) & 47(2) &  76.0(2.8) & 1.57 & 0.80(31) \\ % ok 13.7e-12
\hline 
\textbf{BZB~J1743+1935} &          &    &       &            &            &            &       &            &      &          \\
00030950001             & 07/06/15 & pc &  1918 & 1.97(0.09) & 0.35(0.24) & 1.11(0.30) & 37(2) &  59.8(3.1) & 1.34 & 1.90(19) \\ % ok 8.40e-12
00040639001             & 10/07/10 & pc &   860 & 2.01(0.24) & -          & -          & 36(3) & -          & 1.41 & 0.63(5)  \\ % ok 8.32e-12     4.231e-12/8.36e-12=0.51
sum                     & -        & pc &  2779 & 1.92(0.07) & 0.47(0.19) & 1.22(0.18) & 37(2) &  60.0(2.5) & 1.30 & 1.25(27) \\ % ok 8.27e-12
\hline
\textbf{BZB~J2131-0915} &          &    &       &            &            &            &       &            &      &          \\
00037543001             & 09/03/30 & pc &  5189 & 2.16(0.06) & 0.76(0.22) & 0.78(0.10) & 16(1) &  26.6(1.2) & 4.69 & 0.96(23) \\ % ok 4.20e-12     7.302e-12/4.20e-12=1.74
\hline
\textbf{BZB~J2201-1707} &          &    &       &            &            &            &       &            &      &          \\
00036814001             & 07/12/08 & pc &  4538 & 1.52(0.17) & 0.92(0.42) & 1.82(0.32) &  7(1) &  12.8(1.0) & 0.28 & 0.65(7)  \\ % ok 1.78e-12 
00036814002             & 07/12/11 & pc &  5991 & 2.09(0.12) & -          & -          &  7(1) & -          & 0.26 & 0.64(10) \\ % ok 1.97e-12     2.368e-12/1.875e-12=1.26
sum                     & -        & pc & 10530 & 1.89(0.08) & 0.40(0.18) & 1.36(0.23) &7.0(0.3)& 11.4(0.6) & 0.28 & 0.67(21) \\ % ok 1.95e-12
\noalign{\smallskip}
\hline
\end{tabular}\\
Col. (7) $E_p$ is in keV. \\
Col. (8) $K$ is in 10$^{-4}$~photons~cm$^{-2}$~s$^{-1}$~keV$^{-1}$.\\
Col. (9) $S_p$ is in units of 10$^{-13}$ erg~cm$^{-2}$~s$^{-1}$.\\
Col. (10) $F_X$ denoting the 0.5-10 keV flux measured in units of 10$^{-11}$ erg~cm$^{-2}$~s$^{-1}$. 
\end{table*}

\begin{table*}
\caption{\textit{Swift} ~spectral analysis results with the LP model of the UBLs.}
\tiny
\begin{tabular}{|llllccccccc|}
\hline
\noalign{\smallskip}
Obs ID & $Date$ & Frame & Exps & $a$ & $b$ & $E_p$ & $K$ & $S_p$ & $F_X$ & $\chi^2_{r}$\\ % flux swift 0.1-2.4 e ratio Frosat/Fswift nella stessa banda    <a_X> e <Gamma_G>
\hline
\noalign{\smallskip}
\hline
\textbf{BZB~J2250+3824} &          &    &       &            &            &            &       &            &      &          \\
00039211001             & 09/08/10 & pc &  1110 & 2.34(0.31) & -          & -          & 27(2) & -          & 0.84 & 1.18(4)  \\ % ok 5.61e-12
00039211002	            & 10/02/18 & pc &  2996 & 2.47(0.14) & -          & -          & 15(1) & -          & 0.42 & 1.27(8)  \\ % ok 3.05e-12
00040151001             & 10/04/17 & pc &  1654 & 2.97(0.82) & -          & -          & 15(2) & -          & 0.34 & 1.25(3)  \\ % ok 3.34e-12
00039211003	            & 10/04/18 & pc &  3446 & 2.55(0.12) & -          & -          & 17(1) & -          & 0.38 & 1.11(11) \\ % ok 3.31e-12
sum                     & -        & pc &  9205 & 2.43(0.06) & 0.29(0.18) & 0.19(0.02) & 17(1) &  39.4(12.3)& 0.45 & 0.66(33) \\ % ok 3.43e-12
\noalign{\smallskip}
00039211005	            & 10/10/05 & pc &  1741 & 2.01(0.11) & 1.19(0.26) & 0.99(0.11) & 52(2) &  84.1(3.9) & 1.24 & 0.72(18) \\ % ok 9.61e-12
00039211006	            & 10/10/06 & pc &  4130 & 2.19(0.08) & 0.98(0.25) & 0.80(0.10) & 70(3) & 114.1(5.1) & 1.62 & 0.55(23) \\ % ok 12.9e-12
00039211007	            & 10/10/07 & pc &  3135 & 2.38(0.07) & -          & -          & 58(2) & -          & 1.75 & 1.09(17) \\ % ok 12.0e-12
00039211008             & 10/10/08 & pc &   782 & -          & -          & -          & -     & -          & -    & -        \\ % ok
00039211009             & 10/10/09 & pc &   143 & -          & -          & -          & -     & -          & -    & -        \\ % ok
sum                     & -        & pc &  9932 & 2.17(0.03) & 0.81(0.08) & 0.78(0.05) & 62(1) & 100.8(1.9) & 1.52 & 1.16(111)\\ % ok 11.6e-12
\noalign{\smallskip}
00039211010             & 10/10/10 & pc &  1429 & 2.07(0.13) & 0.87(0.46) & 0.91(0.20) & 49(3) &  78.0(4.8) & 1.24 & 1.20(12) \\ % ok 9.16e-12
00039211011             & 10/10/11 & pc &  3655 & 2.05(0.07) & 0.52(0.19) & 0.90(0.17) & 49(2) &  79.1(2.9) & 1.47 & 1.02(33) \\ % ok 9.69e-12
sum                     & -        & pc &  5084 & 2.07(0.06) & 0.63(0.14) & 0.88(0.11) & 49(1) &  79.5(2.4) & 1.38 & 1.06(47) \\ % ok 9.57e-12
\noalign{\smallskip}
00039211012             & 10/10/12 & pc &  4933 & 2.04(0.09) & 0.66(0.21) & 0.93(0.16) & 62(3) & 100.2(4.2) & 1.76 & 1.22(25) \\ % ok 12.2e-12
00039211013             & 10/10/13 & pc &  4382 & 1.80(0.09) & 0.79(0.20) & 1.33(0.12) & 69(3) & 114.4(4.9) & 2.18 & 0.68(26) \\ % ok 13.6e-12
sum                     & -        & pc &  9316 & 1.91(0.06) & 0.78(0.13) & 1.14(0.08) & 64(2) & 103.3(3.0) & 1.87 & 0.96(55) \\ % ok 12.4e-12
\noalign{\smallskip}
00039211014             & 10/10/14 & pc &  4676 & 1.91(0.09) & 0.87(0.25) & 1.12(0.11) & 63(3) & 100.7(4.3) & 1.76 & 1.26(27) \\ % ok 12.0e-12
00039211015             & 10/10/15 & pc &  5316 & 2.02(0.05) & 1.01(0.14) & 0.97(0.06) & 52(1) &  82.8(2.4) & 1.29 & 1.11(53) \\ % ok 9.64e-12
00039211016             & 10/10/16 & pc &  3304 & 2.02(0.07) & 0.84(0.19) & 0.97(0.10) & 56(2) &  89.2(3.4) & 1.48 & 0.95(31) \\ % ok 10.6e-12     2.449e-12/9.05e-12=0.27
sum                     & -        & pc & 13300 & 1.91(0.05) & 1.22(0.13) & 1.09(0.05) & 57(1) &  91.6(2.4) & 1.41 & 0.93(64) \\ % ok 10.5e-12
\hline
\textbf{BZB~J2308-2219} &          &    &       &            &            &            &       &            &      &          \\
00036815001             & 07/09/29 & pc &  7289 & 1.95(0.20) & -          & -          & 3(1)  & -          & 0.13  & 0.74(3) \\ % ok 0.89e-12     4.27e-12/0.89e-12=4.80
\hline
\textbf{BZB~J2322+3436} &          &    &       &            &            &            &       &            &      &          \\
00040684001             & 10/02/17 & pc &  1344 & -          & -          & -          & -     & -          & -    & -        \\ % ok
00040684002             & 10/02/17 & pc &  3523 & 2.28(0.22) & -          & -          & 4(1)  & -          & 0.14 & 1.70(2)  \\ % ok 1.03e-12     1.091e-12/1.03e-12=1.06
sum                     & -        & pc &  4866 & 2.16(0.30) & -          & -          & 5(1)  & -          & 0.14 & 1.26(3)  \\ % ok 1.09e-12
\hline
\textbf{BZB~J2343+3439} &          &    &       &            &            &            &       &            &      &          \\
00037545001	            & 08/05/30 & pc &  2381 & 1.60(0.13) & 1.34(0.30) & 1.40(0.10) & 24(1) &  41.7(2.5) & 0.73 & 1.17(14) \\ % ok 5.15e-12
00037545002	            & 08/06/02 & pc &  3336 & 1.78(0.09) & 0.82(0.22) & 1.35(0.13) & 24(1) &  39.3(1.9) & 0.79 & 0.90(20) \\ % ok 5.24e-12     5.43e-12/5.195e-12=1.04
sum                     & -        & pc &  5717 & 1.69(0.07) & 0.99(0.17) & 1.43(0.08) & 24(1) &  39.8(1.5) & 0.77 & 1.14(36) \\ % ok 5.12e-12
\noalign{\smallskip}
\hline
\end{tabular}\\
Col. (7) $E_p$ is in keV. \\
Col. (8) $K$ is in 10$^{-4}$~photons~cm$^{-2}$~s$^{-1}$~keV$^{-1}$.\\
Col. (9) $S_p$ is in units of 10$^{-13}$ erg~cm$^{-2}$~s$^{-1}$.\\
Col. (10)$F_X$ denoting the 0.5-10 keV flux measured in units of 10$^{-11}$ erg~cm$^{-2}$~s$^{-1}$. 
\end{table*}

\begin{table*}
\caption{\textit{XMM-Newton} ~spectral analysis results with the LP model of the UBLs.}
\tiny
\begin{tabular}{|llllccccccc|}
\hline
\noalign{\smallskip}
Obs ID & $Date$ & Frame & Exps & $a$ & $b$ & $E_p$ & $K$ & $S_p$ & $F_X$ & $\chi^2_{r}$\\
\hline
\noalign{\smallskip}
\hline
\textbf{BZB~J0208+3523} &          &           &       &            &            &            &       &           &      &          \\
0084140101              & 01/02/14 & M1-FW(Me) & 38070 & 2.09(0.03) & 0.61(0.07) & 0.85(0.06) &8.1(0.1)&13.0(0.2) & 0.24 & 0.92(134)\\ % ok 1.84e-12
0084140501              & 02/02/04 & M1-FW(Me) & 11680 & 1.95(0.05) & 0.41(0.11) & 1.16(0.14) &8.7(0.2)&13.9(0.3) & 0.31 & 1.13(69) \\ % ok 2.03e-12
\hline
\textbf{BZB~J0326+0225} &          &           &       &            &            &            &       &           &      &          \\
0094382501              & 02/02/05 & M1-PW(Me) &  4563 & 2.36(0.07) & 0.32(0.16) & 0.27(0.21) &18.4(0.4)&36.6(7.3)& 0.50 & 0.68(47) \\ % ok 4.00e-12
\hline
\textbf{BZB~J0441+1504} &          &           &       &            &            &            &       &           &      &          \\
0203160101              & 03/09/05 & M1-PW(Th) &  7438 & 2.18(0.17) & -          & -          &3.3(0.1)& -        & 0.11 & 1.57(15) \\ % ok 0.61e-12
\hline
\textbf{BZB~J0744+7433} &          &           &       &            &            &            &       &           &      &          \\
0123100101              & 00/04/13 & M1-PW(Th) & 10620 & 2.17(0.03) & 0.16(0.06) & 0.28(0.18) &25.6(0.3)&45.8(3.6)& 0.94 & 1.01(135)\\ % ok 7.50e-12
0123100201              & 00/04/12 & M1-PW(Th) & 19580 & 2.19(0.03) & 0.18(0.06) & 0.30(0.15) &23.4(0.3)&42.0(2.8)& 0.84 & 1.00(141)\\ % ok 0.68e-12
\hline
\textbf{BZB~J1231+6414} &          &           &       &            &            &            &       &           &      &          \\
0124900101              & 00/05/21 & M1-FW(Th) & 16220 & 2.15(0.04) & 0.25(0.08) & 0.50(0.18) &9.4(0.1)&15.9(0.7) & 0.34 & 1.05(105)\\ % ok 2.07e-12
\hline
\textbf{BZB~J1237+6258} &          &           &       &            &            &            &       &           &      &          \\
0604830201              & 09/07/09 & M2-FF(Th) &  9847 & 1.83(0.09) & 0.60(0.19) & 1.40(0.14) &5.4(0.2)& 9.0(0.3) & 0.21 & 0.85(41) \\ % ok 1.68e-12
\hline
\textbf{BZB~J1257+2412} &          &           &       &            &            &            &       &           &      &          \\
0094383201              & 02/12/12 & M2-SW(Md) &  5788 & 2.00(0.04) & -          & -          &20.9(0.4)& -       & 0.98 & 1.06(95) \\ % ok 7.42e-12
\hline
\textbf{BZB~J1510+333}  &          &           &       &            &            &            &       &           &      &          \\
0303930101              & 06/01/02 & M2-FF(Th) &  9844 & 1.64(0.02) & 0.45(0.11) & 2.53(0.35) &7.8(0.2)&14.9(0.4) & 0.38 & 1.14(64) \\ % ok 2.32e-12
\hline
\textbf{BZB~J1626+3513} &          &           &       &            &            &            &       &           &      &          \\
0505010501              & 07/08/17 & M2-FF(Md) &  7696 & 2.50(0.15) & -          & -          &3.4(0.1)& -        & 0.10 & 0.65(12) \\
0505011201              & 07/08/19 & M2-FF(Md) & 15300 & 2.50(0.09) & -          & -          &3.5(0.1)& -        & 0.11 & 0.93(28) \\ % ok 1.42e-12     0.837e-12/1.42e-12=0.59
\noalign{\smallskip}
\hline
\end{tabular}\\
Col. (7) $E_p$ is in keV. \\
Col. (8) $K$ is in 10$^{-4}$~photons~cm$^{-2}$~s$^{-1}$~keV$^{-1}$.\\
Col. (9) $S_p$ is in units of 10$^{-13}$ erg~cm$^{-2}$~s$^{-1}$.\\
Col. (10)$F_X$ denoting the 0.5-10 keV flux measured in units of 10$^{-11}$ erg~cm$^{-2}$~s$^{-1}$. 
\end{table*}

\begin{table*}
\caption{UBLs selected.}\label{tabella8}
\begin{center}
\tiny
\begin{tabular}{|lccc|}
\hline
BZCAT~Name & $F_{ROSAT}$                     & $<F_{0.1-2.4keV}>$            & $\rho$  \\
\noalign{\smallskip}
\hline
\noalign{\smallskip}
BZB~J0013-1854    & 6.488  & 8.7675  &  0.74 \\
BZB~J0123+3420    & 25.254 & 19.6538 &  1.28 \\
BZB~J0201+0034    & 3.526  & 3.26    &  1.08 \\
BZB~J0208+3523    & 2.879  & 1.935   &  1.49 \\
BZB~J0214+5144    & 4.579  & 8.449   &  0.54 \\
BZB~J0227+0202    & 18.184 & 7.03    &  2.59 \\
BZB~J0325-1646    & 27.164 & 1.48    & 18.35 \\
BZB~J0326+0225    & 12.038 & 1.36    &  8.85 \\
BZB~J0441+1504    & 10.195 & 3.62    &  2.82 \\
BZB~J0442-0018    & 1.673  & 0.55    &  3.04 \\
BZB~J0621-3411    & 1.838  & 4.34    &  0.43 \\
BZB~J0744+7433    & 6.291  & 4.09    &  0.65 \\
BZB~J0751+1730    & 1.781  & 1.53    &  1.16 \\
BZB~J0753+2921    & 1.287  & -       &  -    \\
BZB~J0847+1133    & 11.006 & 9.39    &  1.17 \\
BZB~J0916+5238    & 3.833  & 2.98    &  1.29 \\
BZB~J0930+4950    & 16.672 & 8.85    &  1.88 \\
BZB~J0952+7502    & 2.485  & 2.09    &  1.18 \\
BZB~J1010-3119    & 10.149 & 8.794   &  1.15 \\
BZB~J1022+5124    & 3.442  & 2.75    &  1.25 \\
BZB~J1053+4929    & 0.821  & 1.86    &  0.44 \\
BZB~J1056+0252    & 6.903  & 5.07    &  1.36 \\
BZB~J1111+3452    & 3.998  & -       &  -    \\
BZB~J1117+2014    & 33.576 & 7.22    &  4.65 \\
BZB~J1136+6737    & 14.752 & 17.02   &  0.87 \\
BZB~J1145-0340    & 4.101  & 2.16    &  1.90 \\
BZB~J1154-0010    & 2.475  & -       &  -    \\
BZB~J1231+6414    & 2.489  & 2.07    &  1.20 \\
BZB~J1237+6258    & 2.284  & 2.549   &  0.90 \\
BZB~J1253-3931    & 7.046  & 4.59    &  1.54 \\
BZB~J1257+2412    & 7.221  & 5.35    &  1.35 \\
BZB~J1341+3959    & 5.152  & 5.03    &  1.02 \\
BZB~J1417+2543    & 15.304 & 18.40   &  0.83 \\
BZB~J1439+3932    & 11.076 & 7.325   &  0.66 \\
BZB~J1442+1200    & 7.82   & 7.324   &  1.07 \\
BZB~J1510+3335    & 2.525  & 2.32    &  1.09 \\
BZB~J1534+3715    & 0.228  & 0.69    &  0.33 \\
BZB~J1605+5421    & 3.874  & 1.45    &  2.67 \\
BZB~J1626+3513    & 0.837  & 1.42    &  0.59 \\
BZB~J1728+5013    & 20.374 & 13.1    &  1.55 \\
BZB~J1743+1935    & 4.231  & 8.36    &  0.51 \\
BZB~J2131-0915    & 7.302  & 4.20    &  1.74 \\
BZB~J2201-1707    & 2.368  & 1.875   &  1.26 \\
BZB~J2250+3824    & 2.449  & 9.05    &  0.27 \\
BZB~J2308-2219    & 4.27   & 0.89    &  4.80 \\
BZB~J2322+3436    & 1.091  & 1.03    &  1.06 \\
BZB~J2343+3439    & 5.43   & 5.195   &  1.04 \\
\noalign{\smallskip}
\hline
\end{tabular}\\
$F_{ROSAT}$ is in units of $10^{-12} erg~cm^{-2}~s^{-1}$.\\
$<F_{0.1-2.4keV}>$ is in units of $10^{-12} erg~cm^{-2}~s^{-1}$.\\
\end{center}
\end{table*}

\end{document}